\documentclass[fleqn,12pt]{wlscirep}
\usepackage[utf8]{inputenc}
\usepackage[T1]{fontenc}
\usepackage{xcolor}
\usepackage[font=scriptsize]{caption}
\usepackage{mhchem}
\usepackage{pifont}
\usepackage{chemformula}
\usepackage{filecontents}
\usepackage[left]{lineno}

\newcommand{\cmark}{\textcolor{green}{\ding{51}}}%
\newcommand{\xmark}{\textcolor{red}{\ding{55}}}%
\title{An Ice Age JWST inventory of dense molecular cloud ices}

\author[1,*]{M. K. McClure}
\author[2]{W. R. M. Rocha}
\author[3]{K. M. Pontoppidan}
\author[1]{N. Crouzet}

\author[4]{L.~E.~U. Chu}
\author[5]{E. Dartois}
\author[6,1]{T. Lamberts}
\author[7]{J.~A. Noble}
\author[8]{Y.~J. Pendleton}
\author[9]{G. Perotti}
\author[10]{D. Qasim}
\author[2]{M.G. Rachid}
\author[11]{Z.L. Smith}
\author[12]{F. Sun}

\author[3]{Tracy L Beck}
\author[13]{A. C. A. Boogert}
\author[14]{W. A. Brown}
\author[15]{P. Caselli}
\author[16]{S.B. Charnley}
\author[17]{Herma M. Cuppen}
\author[11]{H. Dickinson}
\author[18]{M.~N. Drozdovskaya}
\author[12]{E. Egami}
\author[13]{J. Erkal}
\author[11]{H. Fraser}
\author[19]{R.~T. Garrod}
\author[20]{D. Harsono}
\author[21]{S. Ioppolo}
\author[22]{I. Jim\'enez-Serra}
\author[16,23]{M. Jin}
\author[24]{J. K. J{\o}rgensen}
\author[24]{L. E. Kristensen}
\author[25]{D.C. Lis}
\author[26]{M. R. S. McCoustra}
\author[27,28]{Brett A. McGuire}
\author[29]{G.J. Melnick}
\author[29]{Karin I. \"Oberg}
\author[30]{M.~E. Palumbo}
\author[31]{T. Shimonishi}
\author[1]{J.A. Sturm}
\author[1]{E.F. van Dishoeck}

\author[2]{H.~Linnartz}

\affil[1]{Leiden Observatory, Leiden University, PO Box 9513, NL--2300 RA Leiden, The Netherlands}
\affil[*]{mcclure@strw.leidenuniv.nl}
\affil[2]{Laboratory for Astrophysics, Leiden Observatory, Leiden University, PO Box 9513, NL--2300 RA Leiden, The Netherlands}
\affil[3]{Space Telescope Science Institute, 3700 San Martin Drive, Baltimore, MD 21218, USA}
\affil[4]{NASA Postdoctoral Program Fellow, NASA Ames Research Center, M/S 245-1, Moffett Field, CA 94035, USA}
\affil[5]{Institut des Sciences Mol\'eculaires d’Orsay, CNRS, Univ. Paris-Saclay, 91405 Orsay, France}
\affil[6]{Leiden Institute of Chemistry, Gorlaeus Laboratories, Leiden University, PO Box 9502, 2300 RA Leiden, The Netherlands}
\affil[7]{Physique des Interactions Ioniques et Mol\'{e}culaires, CNRS, Aix Marseille Univ., 13397 Marseille, France}
\affil[8]{Department of Physics, University of Central Florida, Orlando, FL 32816, USA}
\affil[9]{Max Planck Institute for Astronomy, K{\"o}nigstuhl 17, D-69117 Heidelberg, Germany}
\affil[10]{Southwest Research Institute, San Antonio, TX 78238, USA}
\affil[11]{School of Physical Sciences, The Open University, Walton Hall, Milton Keynes MK7 6AA, United Kingdom}
\affil[12]{Steward Observatory, University of Arizona, 933 N. Cherry Avenue, Tucson, AZ 85721, USA}
\affil[13]{Institute for Astronomy, University of Hawai'i at Manoa, 2680 Woodlawn Drive, Honolulu, HI 96822, USA}
\affil[14]{Department of Chemistry, University of Sussex,  Falmer, Brighton BN1 9QG, UK}
\affil[15]{Max-Planck-Institut f\"ur extraterrestrische Physik, Gie{\ss}enbachstrasse 1, 85748 Garching bei M\"unchen, Germany}
\affil[16]{Solar System Exploration Division, NASA Goddard Space Flight Center, Greenbelt, MD 20771, USA}
\affil[17]{Radboud University, Institute for Molecules and Materials, Nijmegen 6525 AJ, The Netherlands}
\affil[18]{Center for Space and Habitability, Universit{\"a}t Bern, Gesellschaftsstrasse 6, 3012 Bern, Switzerland}
\affil[19]{Departments of Astronomy \& Chemistry, University of Virginia, Charlottesville, VA 22904, USA}
\affil[20]{Institute of Astronomy, Department of Physics, National Tsing Hua University, Hsinchu, Taiwan}
\affil[21]{Center for Interstellar Catalysis, Department of Physics and Astronomy, Aarhus University, Ny Munkegade 120, Aarhus C 8000, Denmark}
\affil[22]{Centro de Astrobiolog\'{\i}a (CAB), CSIC-INTA, Ctra. de Ajalvir km 4, E-28850, Torrej\'on de Ardoz, Spain}
\affil[23]{Department of Physics, Catholic University of America, Washington, DC 20064, USA}
\affil[24]{Niels Bohr Institute, University of Copenhagen, {\O}ster Voldgade 5--7, DK~1350 Copenhagen K., Denmark}
\affil[25]{Jet Propulsion Laboratory, California Institute of Technology, 4800 Oak Grove Drive, Pasadena, CA 91109, USA}
\affil[26]{Institute of Chemical Sciences, Heriot-Watt University, Edinburgh EH14 4AS, Scotland}
\affil[27]{Department of Chemistry, Massachusetts Institute of Technology, Cambridge, MA 02139, USA}
\affil[28]{National Radio Astronomy Observatory, Charlottesville, VA 22903, USA}
\affil[29]{Center for Astrophysics \textbar\ Harvard \& Smithsonian, 60 Garden St., Cambridge, MA 02138, USA}
\affil[30]{INAF - Osservatorio Astrofisico di Catania, via Santa Sofia 78, 95123 Catania, Italy}
\affil[31]{Faculty of Science, Niigata University, Ikarashi-ninocho 8050, Nishi-ku, Niigata, 950-2181, Japan}


\begin{abstract}
\bf {Icy grain mantles are the main reservoir of the volatile elements that link chemical processes in dark, interstellar clouds with the formation of planets and composition of their atmospheres. The initial ice composition is set in the cold, dense parts of molecular clouds, prior to the onset of star formation. With the exquisite sensitivity of JWST, this critical stage of ice evolution is now accessible for detailed study. Here we show the first results of the Early Release Science program ``Ice Age'' that reveal the rich composition of these dense cloud ices. Weak ices, including, $^{13}$CO$_2$, OCN$^-$, $^{13}$CO, OCS, and COMs functional groups are now detected along two pre-stellar lines of sight. The $^{12}$CO$_2$ ice profile indicates modest growth of the icy grains. Column densities of the major and minor ice species indicate that ices contribute between 2 and 19\% of the bulk budgets of the key C, O, N, and S elements. Our results suggest that the formation of simple and complex molecules could begin early in a water-ice rich environment.}
\end{abstract}

\begin{document}
\flushbottom
\maketitle

%
\thispagestyle{empty}
In molecular clouds, the volatile elements that make up life as we know it (carbon, hydrogen, oxygen, nitrogen, and sulfur, i.e. CHONS) are locked up in ices on the surfaces of dust grains. Vibrational modes of these molecular ices are observed in absorption against the near- and mid-infrared continuum provided by field stars located behind clouds. Fully-resolved absorption bands have logarithmic depths directly proportional to the ice column density along the line of sight, allowing model-independent assessment of relative ice abundances within the same beam. At low extinctions in the outer regions of clouds, a mixture of water (H$_2$O), methane (CH$_4$), and ammonia (NH$_3$) ice forms initially through accretion of atomic H in combination with atomic O \cite{Dulieu2010,Ioppolo2008}, C \cite{qasim2020,Lamberts2022}, and N \cite{Hiraoka:1995,Fedoseev2015a} onto silicate/carbon-rich dust grains. Carbon dioxide (CO$_2$) also forms efficiently in this water ice layer.  In the densest and coldest cloud cores, carbon monoxide (CO) freeze-out forms a CO-dominated ice phase \cite{caselli1999,pontoppidan2006spatial}, where CO$_2$ and other simple ice species continue to form. CO and its reaction products can be hydrogenated to produce methanol (CH$_3$OH)\cite{Watanabe2002} or have a hydrogen atom abstracted\cite{Chuang2016}, and subsequent radical-radical reactions can also lead to the formation of other complex organic molecules (COMs). These simple ices and methanol should provide the feedstock for more complex COMs, such as the biomolecule glycine that is seen in comets \cite{Altwegg2016}, some of which are also capable of forming under pre-stellar core conditions \cite{ioppolo2021non}. Ground-based telescopes and space observatories, like the {\it Infrared Space Observatory (ISO)}\cite{gibb2004interstellar}, {\it Spitzer}\cite{boogert2008c2d}, and {\it Akari}\cite{aikawa2012akari}, have probed ice chemical evolution along sightlines through the envelopes of nascent protostars. However, chemical assays of cloud ice have been limited to regions with visual extinctions below A$_V\sim$50 magnitudes, due to the faintness of field stars seen at larger $\rm{A_V}$ \cite{boogert2015observations,Noble2017}.

Here we report the first observations of pristine cloud ices at $A_V>50$ towards two background stars, NIR38 (11:06:25.57 -77:23:15.87, J2000) and SSTSL2J110621.63-772354.1 (hereafter "J110621", 11:06:21.64 -77:23:54.12, J2000), using the James Webb Space Telescope (JWST). These stars probe dense lines of sight just outside the infalling envelope of a Class 0 protostar, Cha MMS1 \cite{belloche2011a} in the low-mass star forming region Chameleon I (192 pc, \cite{dzib2018_distance}).  Initial calculations of their extinction based on the intrinsic colors of K and G giant stars and mid-infrared photometry suggest values of $A_V\sim60$ and $A_V\sim95$, respectively (see section 3.4 in \cite{jin2022ice}), or N$_{\rm H}$=1.1$\times$10$^{23}$ cm$^{-2}$ and 1.7$\times$10$^{23}$ cm$^{-2}$, respectively. 
The observations presented here were obtained with NIRSpec\cite{nirspec_jakobsen2022} Fixed Slit (FS) mode (R$\sim$2600, 2.7--5.3 $\mu$m), NIRCam\cite{nircam_greene2017} Wide Field Slitless Spectrograph (WFSS) mode (R$\sim$1600, 2.4--5.0 $\mu$m), and MIRI\cite{miri_rieke2015PASP} Low Resolution Spectrograph (LRS) FS mode (R$\sim$100, 5--14 $\mu$m) (see Methods section for more details), in order to cover all five major simple ice species, H$_2$O, CO$_2$, CO, CH$_4$, and NH$_3$, and the simplest COM, CH$_3$OH. 

\begin{figure}[htb!]
\centering
\includegraphics[width=\hsize]{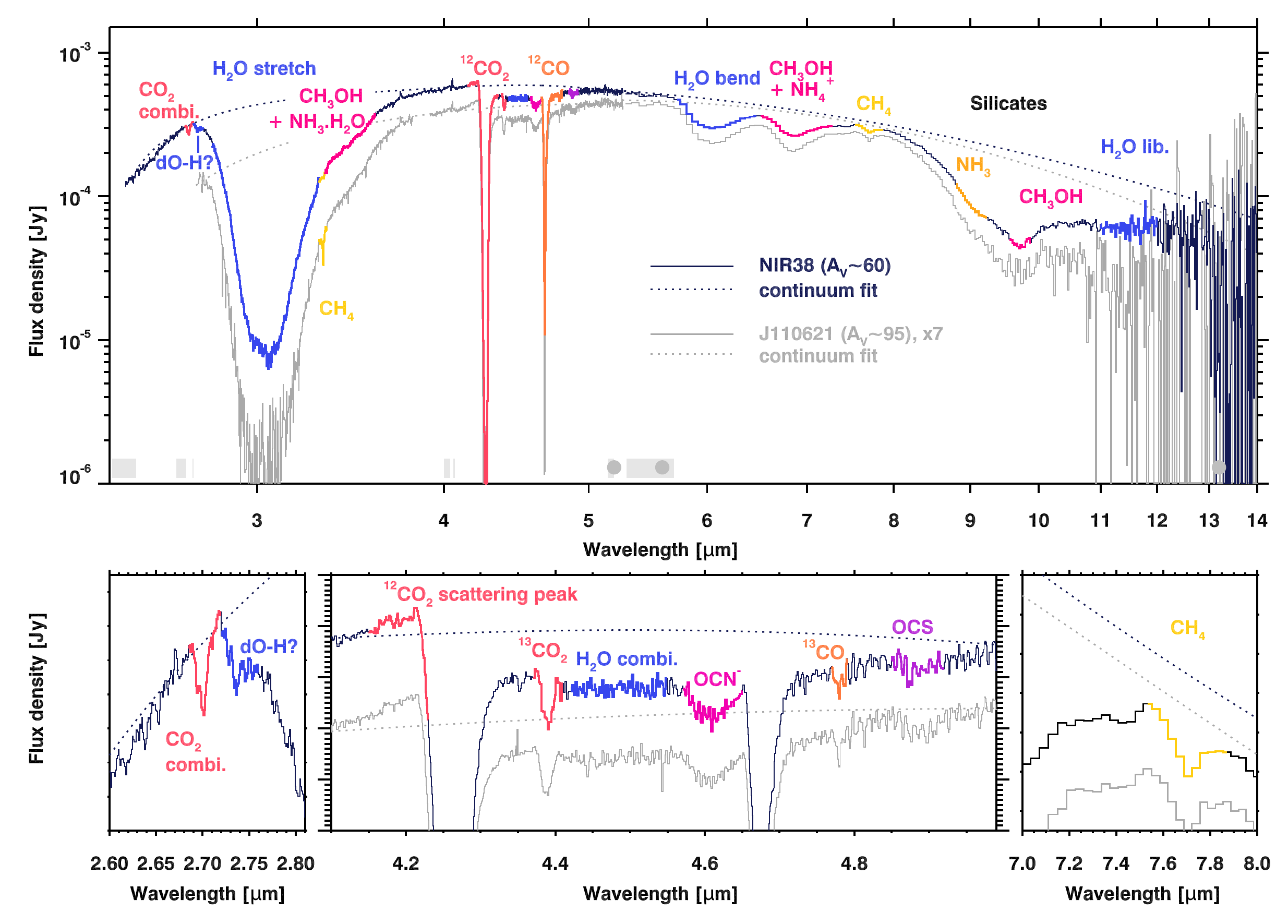}
\caption{\textbf{NIRSpec FS [NIRCam WFSS] and MIRI LRS spectra of NIR38 and J110621.} Top: Full NIRSpec FS and MIRI LRS spectra of NIR38 
($A_V\sim60$, solid navy line) and J110621 ($A_V\sim95$, solid light gray line), with associated continuum fits (dotted lines). For NIR38, a preliminary NIRCam WFSS spectrum has been scaled to the NIRSpec spectrum at 3.8 $\mu$m and spliced in to cover the NIRSpec FS gap from 3.85--3.9 $\mu$m and extend the spectrum to 2.5 $\mu$m. Ice absorption features are color-coded according to species and labelled in the NIR38 spectrum. Wavelength regions used for the continuum fit are indicated by light gray bars (NIRSpec) and dark gray filled circles (MIRI) at the bottom of the top panel. Bottom: Zoom in on the weaker ice features and structure revealed by JWST. The potential dangling O-H feature is indicted by "dO-H", and the combination modes of CO$_2$ and H$_2$O by "combi."}

\label{main_fig1}
\end{figure}

\begin{figure}[htb!]
\centering
\includegraphics[width=\hsize]{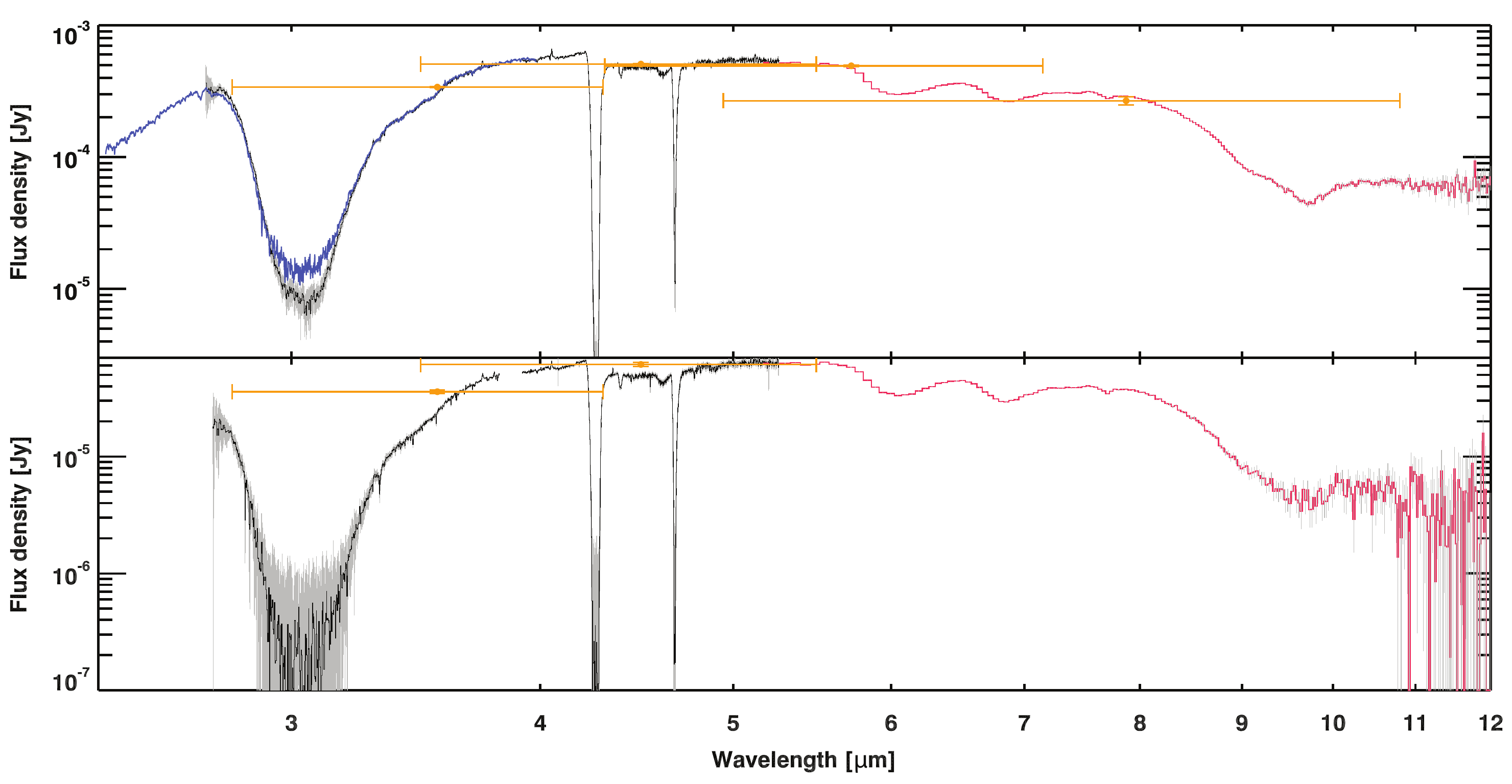}
\caption{\textbf{Data quality comparison for NIR38 and J110621.} (Top panel) Comparison of the NIRCam WFSS (blue), NIRSpec FS (black), and MIRI LRS FS (red) spectra of the $A_V=60$ background star. Error bars (gray) are 3$\sigma$, and in some regions are smaller than the thickness of the lines. Spitzer IRAC photometry (gold points) from the IPAC SEIP catalog is given for reference, with error bars and bandpass indicated. (Bottom panel) Comparison of NIRSpec FS and MIRI LRS FS data for the $A_V=95$ star. Colors are the same as in the top panel.}
\label{main_fig2}
\end{figure}

The full, multi-instrument 2.5--13 $\mu$m spectra towards both high-$\rm{A_V}$ background stars are presented in the top panel of Figure \ref{main_fig1}, with major solid-state features labeled. The identifications of features that we detect, tentatively detect, and do not detect are presented in Table \ref{tab1}.  The spectra obtained from each instrument are compared in Figure \ref{main_fig2}; NIR38 is detected in the continuum at 3.97 $\mu$m by NIRSpec FS with 547.1$\pm$2.6 $\mu$Jy (S/N$\sim$207) and NIRCam WFSS with 551.2$\mu$Jy, while its flux at 7.5 $\mu$m with MIRI LRS is 310.7$\pm$0.6 $\mu$Jy (S/N$\sim$499). J110621 was detected at 3.97 $\mu$m by NIRSpec FS with 54.4$\pm$0.4 $\mu$Jy (S/N$\sim$145) and by MIRI LRS at 7.5 $\mu$m with a flux of 39.7$\pm$0.2 $\mu$Jy (S/N$\sim$208). This high sensitivity allows us to detect both the expected strong absorption features of abundant ice species, as well as a number of weak absorption features that are now detectable through quiescent molecular cloud lines of sight (Figure \ref{main_fig1}, bottom, and Figure \ref{main_fig3}). For these spectra we fit a global continuum to specific continuum regions (see Methods and Figure \ref{main_fig1} caption) to calculate optical depths for these ices.

We report column densities and abundances relative to water of the different ice species in Figure \ref{main_fig4} and Table \ref{tab2}, as determined from both global and local fitting of laboratory data (Table \ref{tab3}) to the optical depths over the whole wavelength range (see Methods and fits given by Extended Data Figures 1-3) and to individual ice features (see Supplementary Data Figures). We also consider the shape of the ice bands, which depends on the local environment, in particular whether the ice is mixed with water or not.

\section*{Results}
\noindent \textbf{Ice inventory and new features - } Both spectra in Figure \ref{main_fig1} display all of the deep features that we expect to be associated with the main icy grain constituents: H$_2$O ice, the main isotopolog of both major C-bearing ices, $^{12}$CO$_2$ and $^{12}$CO, and rocky silicates. The column density of water ice increases from $N_{H2O}\sim 7\times10^{18}cm^{-2}$ to $N_{H2O}\sim 13\times10^{18}cm^{-2}$, respectively, between NIR38 and J110621, while CO$_2$ and CO are present at 10-20\% and 20-40\% of H$_2$O ice. Additionally, the sensitivity and spectral resolution of NIRSpec also allow us to detect a number of new features that probe the structure of these main ices, as well as the chemical diversity of additional small molecules in the ice. 

\underline{Inorganic O- and C-ices - } In these simple ice species, we see structure in the $^{12}$CO$_2$ stretching feature at 4.27 $\mu$m, with both an excess emission over the continuum in the blue wing at 4.2 $\mu$m and a strong absorbing red wing that extends to at least 4.35 $\mu$m. While the continuum shape may change slightly with future photospheric model fits, there is no physically motivated fit that could locally change the continuum enough to erase the warped profile. A similar asymmetric profile is theoretically expected to result from ice mantle growth \cite{dartois2022influence}. An analogous scattering profile is tentatively seen in the CO band at 4.7 $\mu$m, where there is red-shifted absorption below the continuum. However, the blue-shifted CO excess requires confirmation, as it overlaps other absorption features. We also detect both the combination mode of $^{12}$CO$_2$ at 2.7 $\mu$m and perhaps the dangling O-H mode of H$_2$O at 2.74 $\mu$m (see Extended Data Figure 1, panel b inset), the latter of which would signify that some fraction of the water ice is porous or mixed with other species. The $^{13}$C isotopologs of CO$_2$ and CO are both detected (see Extended Data Figures 4 and 5, respectively), superimposed over the 4.6--5.0 $\mu$m CO ro-vibrational gas phase lines originating in the stellar photospheres of these background stars. The $^{12}$CO$_2$/$^{13}$CO$_2$ ratio ranges from 69-87 towards these two lines of sight, while the $^{12}$CO/$^{13}$CO ratio ranges from 99-169.

\underline{N-rich ices - } We detect the main N-carrying ice, NH$_3$, in isolation at 9.1 $\mu$m after the removal of the broad 10$\mu$m silicate feature profile from the optical depth spectrum (see Extended Data Figure 6), along with a blended ammonium (NH$_4^+$) feature at 6.85 $\mu$m, which have both been seen before towards dense cores. However, with JWST we are now able to detect the cyanate anion (OCN$^-$) at 4.62 $\mu$m, where it overlaps with the blue scattering wing of $^{12}$CO (Extended Data Figure 7). Ammonium (NH$_4^+$), a potential counter ion, is also detected, securing the identification of OCN$^-$. In contrast, we do not detect other small nitriles, such as CN, HCN, and \ce{CH3CN} (see Methods). The upper limits to these ice column densities range from 0.7 to 2\% of H$_2$O in our spectra. This limit is similar to the 0.1-1\% level of HCN seen in comets \cite{MummaCharnley2011}.

\underline{S-rich ices - } In these spectra, we detect the S-bearing ice species carbonyl sulfide (OCS) around 4.9 $\mu$m, superimposed on the stellar photospheric CO absorption features (Extended Data Figure 8). The simultaneous detection of OCS and CO ice is consistent with a solid-state formation mechanism of CO + S $\rightarrow$ OCS \cite{ferrante2008_ocs}, but constraining the intimate chemical environment of OCS would require careful removal of the photospheric features. There are hints of another S-bearing ice, SO$_2$, at 7.6 $\mu$m in the blue shoulder of the CH$_4$ feature, with detection limits of 0.1-0.3\% with respect to water. The source of sulfur for OCS and potentially SO$_2$ could be from gas-phase depletion into the ice \cite{laas2019modeling}, as well as from minerals, such as troilite (FeS)\cite{kohler2014hidden}. However, the dominant S-bearing ice in comets, hydrogen sulfide (H$_{2}$S)\cite{Calmonte2016}, remains undetected at an upper limit of 0.6\% of H$_2$O, as the 3.92 $\mu$m feature is not detected towards NIR38. This limit is comparable to the $1\%$ level of H$_2$S seen in comets \cite{MummaCharnley2011}.

\begin{figure}[htb!]
\centering
\includegraphics[width=\hsize]{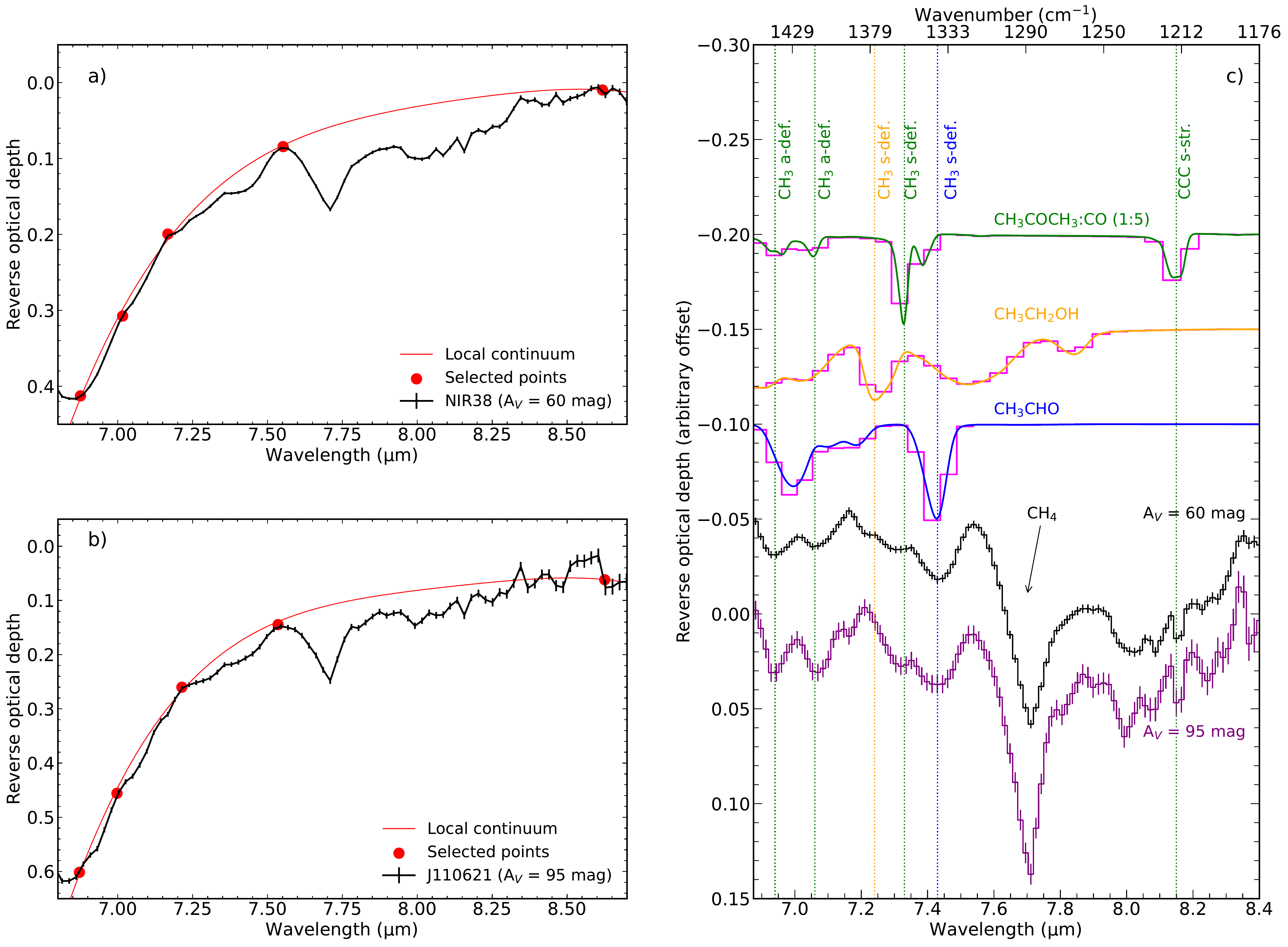}
\caption{Detections of complex organic molecule (COMs) functional groups. Panels (a) and (b): Local continuum over the optical depth spectra of NIR38 ($A_V = 60$~mag) and J110621 ($A_V = 95$~mag) in the range between 6.9 and 8.6~$\mu$m. Panel (c): Local continuum subtracted spectra of NIR38 ($A_V = 60~mag$) and J110621 ($A_V = 95~mag$) compared to laboratory IR spectrum of COMs (CH$_3$CHO - green line\cite{tvs18_comslab}, CH$_3$CH$_2$OH - orange line\cite{tvs18_comslab}, and CH$_3$COCH$_3$:CO - blue line\cite{rachid2020}) in the solid phase. The magenta line shows the laboratory spectra degraded to a resolving power of 150. The vertical lines indicate the match of the experimental data with the observations. The vibrational modes of the experimental data are indicated.}
\label{main_fig3}
\end{figure}

\underline{Organic ices - } We detect both bands of the simple organic ice CH$_4$ for the first time in background stars, at 3.32 and 7.6 $\mu$m. Another low-contrast feature appears from 3.35--3.6 $\mu$m in the red wing of the water ice band. This feature has been detected before towards background stars, but we detect it here with a S/N of 150 and 70 in the A$_V$=60 and 95 sources, respectively. 
At this sensitivity, the feature separates into four distinct peaks that are reproducible between the NIR38 NIRCam and NIRSpec spectra, as well as between NIR38 and J110621 (see Methods). These features are consistent with a blend between the C-H stretch of CH$_3$OH and a broad component centered at 3.47$\mu$m (Extended Data Figure 9).
Ammonia hydrates ({NH$_3$ $\cdot$ H$_2$O}) are considered to be the primary contender for this broad component \cite{dartois2001search}, but the sensitivity of our observations will enable a differential diagnosis in a future work. As seen in previous dense cloud spectra, methanol ice is detected additionally in isolation at 9.7 $\mu$m and blended with the  NH$_4^+$ feature at 6.85 $\mu$m. There is excellent agreement between the column densities derived from both methanol features (See Supplementary Figure 2).
Although the 6 and 6.85 $\mu$m features appear smooth at R$\sim$100 in both sources, 
there are weak but robust absorption excesses at 6.94, 7.06, 7.24, and 7.43~$\mu$m, (see Figure~\ref{main_fig3}), attributable to the functional group in COMs caused by the asymmetric deformation mode of CH$_3$\cite{tvs18_comslab}, 
which has been tentatively detected with {\it Spitzer}\cite{boogert2008c2d} and {\it JWST}\cite{yang2022corinos}. These two bands are seen in the IR spectra of acetone (CH$_3$COCH$_3$)\cite{rachid2020}, ethanol (CH$_3$CH$_2$OH)\cite{tvs18_comslab}, and acetaldehyde (\ce{CH3CHO})\cite{tvs18_comslab}. These background stars will require the higher spectral resolution of MIRI MRS to confirm these identifications, determine the COMs chemical environment, and the degree to which complex chemistry has begun along the J110621 sightline. \\

\begin{figure}[htb!]
\centering
\includegraphics[width=\hsize]{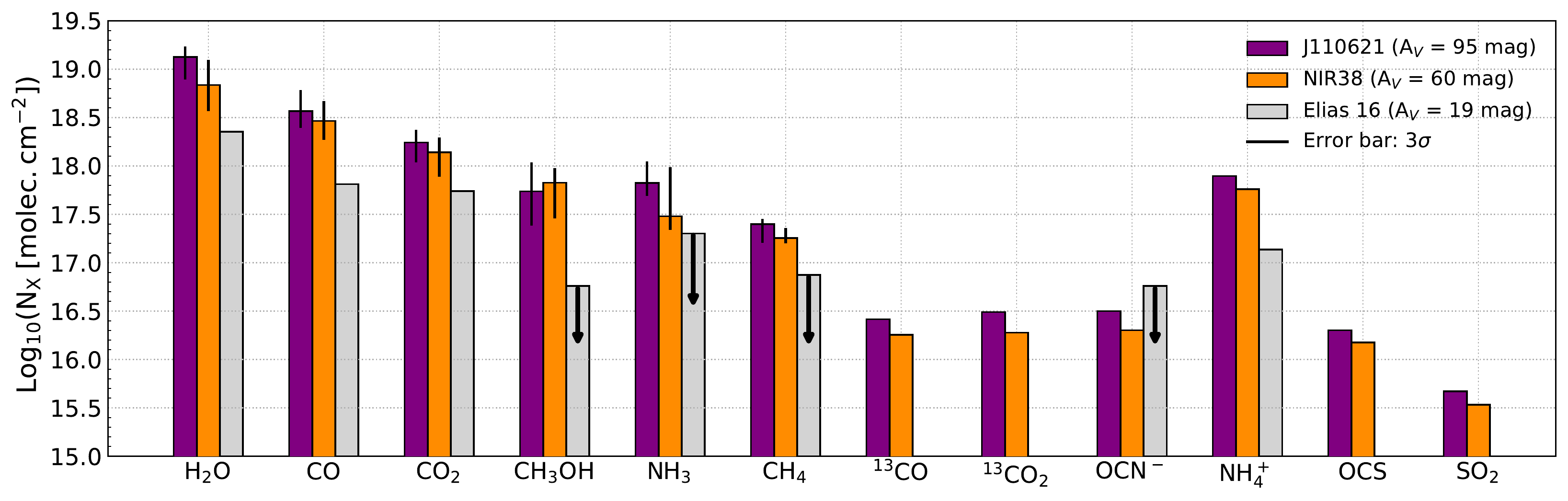}
\includegraphics[width=\hsize]{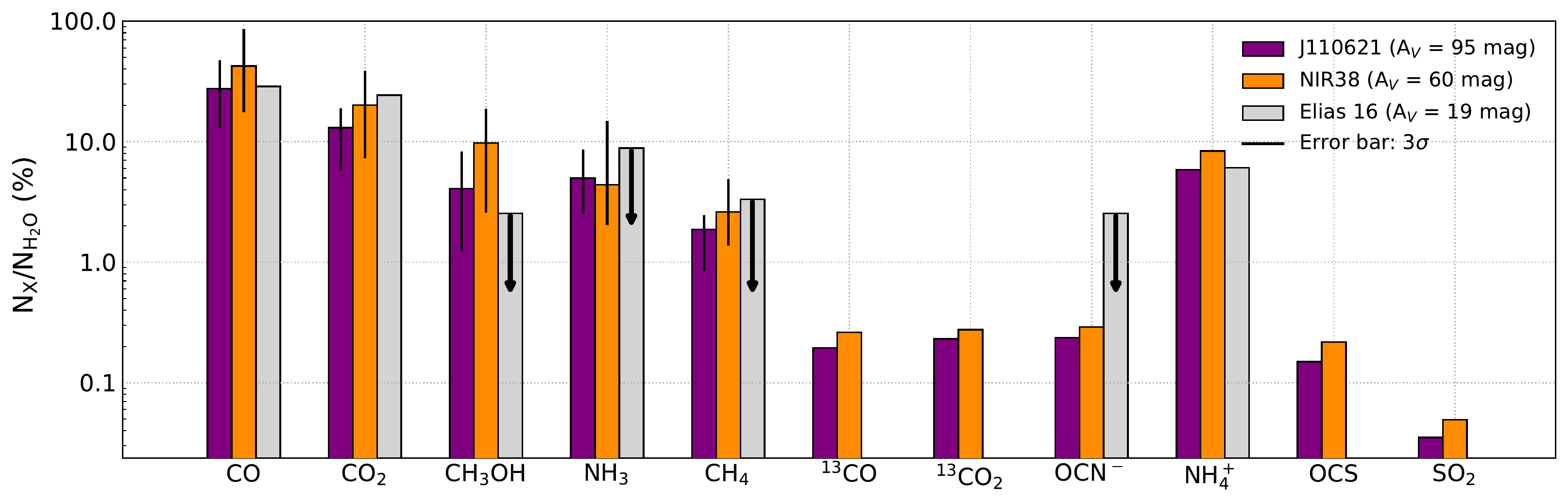}
\caption{\textbf{Barplots showing the derived ice column density for different species  towards NIR38 ($A_V \sim 60$ mag) and J110621 ($A_V \sim 95$ mag).} (Top) Column densities of the ice species identified in this work, compared to the literature values of Elias~16 ($A_V \sim 19$ mag)\cite{Knez2005}. The column densities of the major ice components are from the global \texttt{ENIIGMA} fit (best of n=112 models), and we use the values from the local fits for the minor ice components. Black arrows indicate upper limits and error bars are taken from the 3$\sigma$ confidence intervals. (Bottom) Relative column densities of the detected ices, normalized to H$_2$O ice. For the major ice components, we use the values from the global \texttt{ENIIGMA} fit, and for the minor ice components, we use the values from the local fits. Black arrows indicate upper limits and error bars are taken from the 3$\sigma$ confidence intervals.}
\label{main_fig4}
\end{figure}

\noindent \textbf{Stable ice chemical environment from A$_V\sim$20 to 95 - }
The absolute column densities of most ice species are slightly larger towards J110621, as expected from the increase to $\rm{A_V}\sim$95. However, the ice inventory is very similar towards both sightlines, suggesting that although the total amount of ice increases, the ice composition is set at a lower $\rm{A_V}$. In fact, the relative column densities of the simple ices from 60<A$_V$<95 are broadly consistent with the ice evolution sequence proposed on the basis of {\it Spitzer} observations from 20<A$_V$<50 \cite{boogert2015observations}, as exemplified by the comparison with the background star Elias 16 ($\rm{A_V}\sim$19) in Figure \ref{main_fig4} well as laboratory data and chemical modeling of this dense cloud region \cite{jin2022ice}. These results could suggest that, although CO ice is the second most abundant species detected in our spectra, the local cloud gas density may be less than the limit of $n_H\sim10^{5}cm^{-3}$ required for CO to catastrophically freeze out via collisions \cite{caselli1999,pontoppidan2006spatial}. Supporting this, initial modeling of the $^{12}$CO ice profiles (Extended Data Figures 1 and 2) suggests that they may be dominated by a pure component, with two additional weaker components mixed with methanol or CO$_2$ \cite{boogert2015observations}. In contrast, the local $^{13}$CO$_2$ ice profiles of both stars suggest that CO$_2$ is dominated by an intimate mixture with H$_2$O, with a lesser contribution from a CO-rich mixture (Methods and Extended Data Figure 4). Additionally, based on comparisons between laboratory data and the profiles of the $^{12}$CO and 9.7 $\mu$m methanol bands, methanol seems to reside in environments containing both \ce{H2O} and CO (see Methods, Extended Data Figures 1 and 2). The amount of methanol appears to be approximately the same in both sources, based on both the 3.53 $\mu$m and 9.7 $\mu$m features. In contrast, there are constant or increasing column densities of the simple hydrides, NH$_3$ and CH$_4$.

\section*{Discussion}
The sum of the column densities for both CO isotopolog ices and their potential reaction products is less than the expected total CO column density from $\rm{A_V}$ for each line of sight, suggesting that at most 46\% and 33\% of the available CO gas has frozen out into ices towards NIR38 and J110621, respectively. Although NIR38 samples a smaller total column of dust, its line of sight appears to pass closer to the Class 0 protostar (see Extended Data Figure 10). If this region contains locally denser or colder dust, it could explain the larger fraction of total CO that is frozen out onto the grains. These results imply that the rich variety of ices that we see likely formed early, prior to catastrophic CO freeze-out, rather than later through purely CO hydrogenation pathways. 

The other ice column density results are also consistent with efficient, early formation of CO$_2$, NH$_3$, and CH$_4$ in water-rich ices through H-addition and abstraction \cite{Goumans2008,garrod2011,qasim2020,Fedoseev2015a}, followed by a small amount of the subsequent CO-based chemistry that we would expect to see at these high extinctions. While methanol was traditionally thought to form efficiently via successive CO hydrogenation, with \ce{H2CO} as an intermediate\cite{Watanabe2002}, it can also form earlier and more slowly in the \ce{H2O}-rich ice phase \cite{qasim2018formation,molpeceres2021carbon}. Our fit to the  feature at 9.7 $\mu$m suggests that both formation pathways may operate in these ices. This conclusion is supported by the detection of functional groups of COMs at 7-7.5$\mu$m towards pre-stellar sources that lie outside of the coldest cores in this region, suggesting their early formation in the water rich phase \cite{Chuang2016}. Models predict only ethanol at N$\sim$6-15$\times$10$^{16}$ cm$^{-2}$,\cite{jin2022ice} which is broadly consistent with the optical depths at 7.2 $\mu$m in both spectra, but not the other potential COMs species that could produce the other absorption features seen at 6.94, 7.06, and 7.34 $\mu$m. Our detections of COM functional groups in these pre-stellar ices hint at the non-energetic complexity achieved in ices already before the formation of a hot protostellar core.

Accounting for the amount of C, O, N, and S in the ices is critical to determine the bulk volatile budget of the stellar and planetary systems that will form within this molecular cloud. Comparing the column densities of the detected ices for both NIR38 and J110621 with the expected cosmic abundances for C, O, N, and S, we see at most 19\% of the total O- and C-, 13\% of the total N-budget, and 1\% of the S-budget in this dense cloud (see Methods). These numbers are similar to what has been previously reported for protostars \cite{boogert2015observations}, but now we are able to trace the budgets of these elements back to their initial conditions in dense clouds. Most of the remaining budgets will be made up of refractory species, including silicates and amorphous carbons, or other ices like $N_2$ that do not show spectral features at these wavelengths. Some of the budget may additionally be accounted for in COMs that we cannot yet identify conclusively with the MIRI LRS spectral resolution.  

The profile distortions of the deepest ice bands show that the increase in H$_2$O, CO$_2$, and CO is accompanied by an increase in the size of these icy grains.  The enhanced, red-shifted absorption wing, as seen in H$_2$O and CO$_2$ (e.g. \cite{Brooke1996study}), in addition to the blue-shifted emission wing described earlier for the $^{12}$CO$_2$ and possibly $^{12}$CO ice features \cite{dartois2022influence}, are associated with scattering effects resulting from icy grain growth to sizes on the order of the wave vector at which they are detected, i.e. a few microns. Whereas red wing extinction due to scattering is a rather robust effect produced by larger grains, the intensity of the blue emission excess can be highly variable. Its strength is highly sensitive to specific local changes in the optical constants of the grains' core and mantle materials. The profiles of our observed CO$_2$ ice features imply growth to sizes of around 1 $\mu$m, as predicted by some grain growth models (e.g. \cite{silsbee2020rapid}, \cite{ormel2011dust}). Despite this relatively modest increase in maximum grain size, the observed growth occurs at the expense of the smaller grains, which are depleted. Our observed change in icy dust grain size distribution would not only influence the visual extinction but also reduce the total grain surface available for reactions in such dense regions. However, our tentative detection of the OH dangling mode of H$_2$O near 2.7 $\mu$m could suggest that the water in these large grains is porous or mixed with other ices. In that event, the pore surfaces may also provide space for additional reactions.

Detailed modelling to quantify the maximum grain size, shape, and porosity of these ices will be presented in a future work. Further analysis of the reaction pathways and relative ice abundances requires both chemical modeling and future observations of molecular clouds at both low and high $\rm{A_V}$s to confirm when the simple hydrides are formed in relation to CO freeze-out.  Complementary molecular gas phase observations will also confirm the extent to which CO has frozen out in this region. Such work will in part continue through another component of the Ice Age ERS program, in which we have obtained hundreds of ice spectra in the same Chameleon I region with the new multi-object capabilities of NIRCam WFSS. By combining these datasets, the superlative sensitivity, spectral resolution, and wavelength coverage with JWST now enable us to fully probe the initial conditions of all of the major ices in molecular cloud cores just prior to their collapse to form protostars. These new capabilities open the door to understanding the formation and inheritance of these key CHONS-species through the star- and planet-formation process and, ultimately, address what role they will play in shaping the chemistry on emerging planets. \\

\section*{Methods}
\label{methods}

\textbf{Observations and data reduction}\\

\noindent \textbf{NIRSpec: } NIRSpec Fixed Slit (FS)\cite{nirspec_jakobsen2022} observations of the targets NIR38 (A$_V$ = 60) and SSTSL2J110621.63-772354.1 (A$_V$ = 95) were taken on July 6, 2022 and July 8, 2022, respectively, using the G395H grating, combined with F290LP blocking filter. Target acquisition was achieved using the Wide Aperture Target Acquisition (WATA) method and the SUB2048 subarray with the CLEAR filter and a readout pattern of NRSRAPID6 with an exposure time of 14.5 seconds for both sources. Spectra of the two stars were obtained using the ALLSLITS subarray at four dither positions, spaced along the S200A2 slit. Each integration was composed of 57 groups and 265 groups using the NRSRAPID readout pattern, for total on-source exposure times of 1274.7 seconds and 5845.7 seconds, respectively. 

The JWST calibration pipeline was used for detector level 1 processing to calculate rate files from the uncalibrated ramps using version 1.7.1, Calibration Reference Data System (CRDS) context {\tt jwst\_0948.pmap}, and the PUB CRDS server. The two-dimensional rate spectra were distortion corrected using a second-order trace function derived from a commissioning observation of the standard star TYC 4433-1800-1, observed as part of program PID 1128. The two-dimensional spectral dithers were pairwise differenced to efficiently remove the background, and a one-dimensional spectrum from each dither was then optimally extracted\cite{horne1986optimal} using a cross-dispersion profile calculated by median-collapsing each dither in the spectral direction. An uncalibrated spectrum was then derived using a median for the four separate dithers to remove most cosmic rays. Note that the observation of SSTSL2J110621.63-772354.1 used very long integrations ($>$1000\,s), and suffers from large numbers of cosmic ray hits, not all of which could be fully corrected. To flux calibrate the spectra, we extracted spectra from the identically-processed level 1 rate files of the standard star observation of TYC 4433-1800-1 using the same grating and slit. By dividing with the standard star spectrum, and multiplying by a model spectrum of the standard star, scaled to $K_S=11.584\,$mag, we arrive at the final, calibrated spectra. This yielded excellent results, although the direct use of the standard star leave a small number of artifacts from uncorrected hydrogen absorption lines in the standard star spectrum. Note that this process does not rely on pipeline flat fields or calibrations, which are not yet available. However, the wavelength calibration does use the solution from the pipeline. Errors were formally propagated from pixel errors estimated by the ramps-to-slopes fits from the level 1 processing. \\

\noindent \textbf{NIRCam: } NIRCam Wide-field Slitless Spectroscopic (WFSS)\cite{nircam_greene2017} grism observations of the Cha-MMS 1 field were taken on July 3, 2022 with the F322W2 filter (2.5-4.0 $\mu$m) using Grism C with NIR38 (A$_V$ = 60) residing in module A. We obtained 24 individual integrations of the field with a total exposure time of 1.7 hours. We followed a data reduction routine similar to that in \cite{sun2022first}. We first reduced the grism spectroscopic data with the standard JWST calibration pipeline v1.6.2 to the level of Stage-1, using the default CRDS setup with JWST OPS and no modifications, i.e. CRDS context 0953, and then performed 2D sky-background subtraction using the sigma-clipped median images that were constructed from the obtained WFSS data. Flat-field correction was also applied using the imaging flat data obtained with the same filter. We then extracted the spectra of the two background stars using the optimal extraction method \cite{horne1986optimal} from each individual integration, and co-added them together using the SpectRes package \cite{carnall2017spectres}. The wavelength and flux calibrations were performed using the in-flight measurements obtained with JWST Commissioning Program \#1076. At this stage, it is important to note that the current background subtraction method has not been fully optimised, so small systematic offsets may exist within data. Additionally, the optimal extraction method reduces, but may not entirely eliminate, potential flux contamination from other nearby sources. Therefore we may be marginally overestimating the flux for our A$_V$ = 60 source. \\

\noindent \textbf{MIRI LRS: } MIRI Low Resolution Spectrograph (LRS) Fixed Slit (FS)\cite{miri_rieke2015PASP} observations of the targets NIR38 (A$_V$ = 60) and SSTSL2J110621.63-772354.1 (A$_V$ = 95) were taken on July 4, 2022 and July 11, 2022, respectively. Target acquisition was achieved using the F560W filter with a FAST readout pattern with 4 groups per integration for an exposure time of 11.1 seconds for both sources. These two targets used observations with 40 groups per integration and 104 groups per integration, respectively, with 5 integrations per exposure with a two-nod dither pattern along the slit, for a total of 10 total integrations per source and total exposure times of 1132.2 seconds and 2908.2 seconds, respectively. The FASTR1 readout pattern was used. 

We reduced the data with the same procedure for the two sources. We used the STScI JWST pipeline (https://jwst-pipeline.readthedocs.io) version 1.8.2, the PUB CRDS server, and CRDS context {\tt jwst\_0986.pmap} to obtain the Stage 1 and Stage 2 products. We started from the uncalibrated data (Level1b, `uncal' files). We ran the Detector1Pipeline with default parameters and the Spec2Pipeline step by step. We used each dither position as a background image for the other and subtracted the background pixel-wise. From the calibrated images (`cal' files), we extracted a one-dimensional spectrum from each dither position using the optimal extraction method \cite{horne1986optimal} where a cross-dispersion profile is calculated by median-collapsing the 2D spectral trace in the spectral direction. The spectra from both dither positions were averaged to obtain a final spectrum.
As a comparison, we extracted a spectrum using the JWST Spec3Pipeline. We combined the two dither positions into a single image using the `resample\_spec' step and extracted the 1D spectrum using the `extract\_1d' step. We defined the extraction region in a custom reference file and disabled the offset that accounts for the expected location of the source (`use\_source\_posn' set to `False'). This ensured that the aperture was centered on the source. We also extracted spectra from the `cal' files using a simple aperture method (not relying on `extract\_1d') and from the resampled image using the optimal extraction method. All these spectra are in good agreement but the optimal extraction method applied on the `cal' files provides a smoother spectrum, which we kept for scientific interpretation. The CRDS context {\tt jwst\_0986.pmap} uses a wavelength calibration that has been updated for MIRI LRS Fixed Slit after an initial mismatch that was found between the flight calibration and the first extracted science spectra. This new wavelength calibration (encoded in the `jwst\_miri\_specwcs\_0005.fits' reference file) is in good match with the known spectral features detected in our spectra. \\

\noindent \textbf{Data quality: }
The spectra from all three instruments are shown in Figure \ref{main_fig2} for the $\rm{A_V}$=60 star. The flux calibration of these data is such that they match each other within their respective 3$\sigma$ error bars. The differences in the signal at the bottom of the 3 $\mu$m H$_2$O feature are due to the increased sensitivity of NIRSPEC FS relative to NIRCAM WFSS, but are within the error bars. Both stars are saturated in the 4.3 $\mu$m $^{12}$CO$_2$ ice bands and we lose the signal at the bottom of these features, which occurs as well in the 3 $\mu$m band of the $\rm{A_V}$=95 star. The reproducibility of the spectral features between NIRCam and NIRSpec is excellent, and the spectra are broadly consistent with the {\it Spitzer} IRAC photometry of this source given in the SEIP Source List server (https://irsa.ipac.caltech.edu/cgi-bin/Gator/nph-dd), taking into account the lack of convolution with the IRAC filter and the assumed color correction. \\

\noindent \textbf{Global continuum fit: }
The continuum shape of background stars, which is physically limited to be a stellar Rayleigh Jeans tail with ices superimposed on it, is fitted either with detailed stellar photosphere models or simple piece-wise polynomial continua to each star \cite{boogert2011ice}. We do not yet have photospheric models for these stars, so we use polynomial fits: in our case, one over the NIRSpec range and another over the MIRI range. For NIR38 ($\rm{A_V}$=60) we fit the continuum using a fifth order polynomial with continuum points of 2.4-2.49 $\mu$m, 2.65-2.69 $\mu$m, 2.715-2.720 $\mu$m, 4.0-4.04 $\mu$m, 4.06-4.07 $\mu$m, 5.15-5.2 $\mu$m, 5.3-5.7 $\mu$m. For J110621 ($\rm{A_V}$=95), we fit the following continuum points: 2.74-2.78 $\mu$m, 3.98-4.01 $\mu$m, and 5.4-5.7 $\mu$m.  The continuum determination in the MIRI LRS range is not straightforward due to the broad ice and silicate features. We determine the continuum on the MIRI LRS range (5.2$-$13~$\mu$m) using a second-order polynomial function. We set the continuum points at 5.2, 5.6, and 13.2~$\mu$m for NIR38 and J110621. Then we combine the two continua in a piece-wise fashion, with the cutoff between them taken at 5.1 $\mu$m. The continuum shape may change slightly when the more detail stellar model is applied. However, our initial steps in stellar modeling (not discussed here) show good agreement with the polynomial continuum fit found here. We estimate that the uncertainty introduced by the continuum is within the uncertainty in the A-values used to derive the column densities.

The silicate absorption band is removed by a synthetic silicate spectrum composed of amorphous pyroxene\cite{dorschner1995} (Mg$_{0.7}$Fe$_{0.3}$SiO$_3$) and olivine\cite{dorschner1995} (MgFeSiO$_4$), as previously used in the literature\cite{boogert2011ice} for background stars (see Extended Data Figure 6). We used the optool code\cite{dominik2021} to create a synthetic spectrum assuming grains of 1~$\mu$m. We aim at matching the spectral ranges between 8.3 and 8.7~$\mu$m and between 10.1 and 10.4~$\mu$m. In both sources, the absorption of pyroxene dominates over olivine at 9.8~$\mu$m. For the $A_V = 60$ mag star, pyroxene and olivine corresponds to 60\% and 40\%, respectively, whereas for the the $A_V = 95$ mag star, the absorptions are 70\% due to pyroxene and 30\% due to olivine.\\

\noindent \textbf{ENIIGMA global fitting and local fits: }
We used the \texttt{ENIIGMA} fitting tool\cite{Rocha2021,rocha2022} to simultaneously fit multiple features across the NIRSpec and MIRI/LRS range by scaling laboratory ice spectra to match the optical depths in Figure 1. A full list with data used in this paper is shown in Table~\ref{tab3}. It is worth mentioning that these laboratory data are previously baseline corrected and noise smoothed at relevant bands. No further processing is performed during the fitting procedure. In the fits, we assume saturated bands at 3~$\mu$m and 4.27~$\mu$m because of negative fluxes. At these two bands, the fit is not limited by the peak of the band. This is an important assumption to make to avoid underestimating column densities of the molecules contributing to the absorption of these bands. \texttt{ENIIGMA} searched for the best combination of experimental data measured at temperatures of 15~K or below. Motivated by previous works, we explored combinations with ice mixtures composed of CO:CO$_2$\cite{pontoppidan2008c2d}, CO:CH$_3$OH\cite{Cuppen2011}, H$_2$O:CH$_3$OH\cite{Perotti2020} and H$_2$O:NH$_3$\cite{dartois2001search}, H$_2$O:CO$_2$:CH$_4$\cite{Oberg2008}, and pure CO\cite{pontoppidan2003}. In addition to these data, \texttt{ENIIGMA} tested other IR spectra measured at temperatures below 16~K. We did not include the spectrum of the ammonium ion (NH$_4^+$) in the global fits because it is not a consensus that the 6.85~$\mu$m is attributed to this chemical species. A dedicated study of this spectral feature will be performed in a follow-up paper by considering different chemical environments where NH$_4^+$. Additionally, since NH$_4^+$ is formed by a chemical reaction between other molecules (e.g., NH$_3$, HNCO) induced by temperature (warm-up) or radiation (e.g., ultraviolet, X-rays, cosmic rays), the spectrum shows other products that have to be taken into account when making assignments of the IR bands. Overall, \texttt{ENIIGMA} provides a good global fit of the major ice components in the observations, which are used to derive the ice column densities (see Table~\ref{tab2}). They are calculated by $N_X = \int\tau{_\nu} d\nu / A$, where $\int \tau{_\nu} d\nu$ is the integrated optical depth of a specific band, $A$ is the band strength, and $X$ is the chemical species. The uncertainties are derived from 3$\sigma$ confidence intervals based on correlation plots shown in Extended Data Figure 3. Additional sources of uncertainties are not considered in these values. \texttt{ENIIGMA} does not fit entirely the isotope bands of $^{13}$CO$_2$ and $^{13}$CO at 4.38~$\mu$m and 4.78~$\mu$m, respectively. First, this is because the global fit limits the amount of the isotopes by the strong $^{12}$CO$_2$ and $^{12}$CO bands. Second, the isotope abundances in the gases used to make the ice samples in the laboratory may not be the same as found in these astronomical targets. By performing local fits, the chemistry of the isotope bands is better constrained (see Extended Data Figure 4). Nevertheless, the ice column densities are similar to the values obtained with the global fits as seen in Table~\ref{tab2}. 

Local fits are also used to calculate the ice column densities of the major components (see Figure 1 of the Supplementary Data). For the H$_2$O ice, we scale the pure H$_2$O ice IR spectrum at 15~K\cite{Oberg2007} to match the ranges around 2.85$-$2.95 $\mu$m and 3.17$-$3.23 $\mu$m because of the saturation of the bands. The broadband between 5 and 8~$\mu$m are fitted by NH$_4^+$ and H$_2$O as scaled to the 3~$\mu$m band. Independently of the global fit, the NH$_4^+$ spectrum can be locally scaled to the astronomical data since this method does not take into account the contribution of the chemical specie at other wavelengths. The goal of the local fits is to estimate the highest amount of a specific component to the absorption band including or not blending effects with other molecules. Since the contribution of CH$_3$OH absorption is minimal at 6.85~$\mu$m (see Extended Data Figures 1 and 2), we do not include methanol in the local fit of this band. In the cases of CO$_2$, CO, CH$_4$, SO$_2$, NH$_3$ and CH$_3$OH, we adopted Gaussian profiles to fit the A$_V$=60 mag and A$_V$=95 mag spectra, and calculate the ice column densities. Around 4.67~$\mu$m and 7.7~$\mu$m, we adopted more than one sub-component to fit the observations, following the previous studies of these two bands\cite{pontoppidan2003, Oberg2008}. For CH$_3$OH, we also perform a local fit analysis around 3.5~$\mu$m, and the column densities are similar to both local and global fits at 9.8~$\mu$m. The local column densities are compared with the global column densities in the Supplementary Information to validate the global fits.

The ice column densities derived from the global and local fit are collated in Table \ref{tab2}. In Figure~\ref{main_fig4}, we show the column densities of the major ice species derived from the global fits and the minor species derived from the local fits. Additionally, we show a comparison with the column densities derived for a background star with $A_V = 19$ mag\cite{Knez2005}. These column densities are normalized to H$_2$O ice in the bottom panel of Figure~\ref{main_fig4}. In the Supplementary Information, all the values from global and local fits, and from the ranges reported in the literature are compared. A caveat in the \texttt{ENIIGMA} methodology, is that it does not perform grain shape correction of the ice bands. Such a correction comes with a level of discussion beyond the scope of this paper, for example, which grain shape better reproduces the observations, and what are the size distributions. These geometry effects will be explored in a subsequent study. \\

\noindent \textbf{Local continuum fit for weak features:}
To separate the weaker features from the wing of the water stretch and combination bands, we also fit a local continuum to both spectra. The continuum points were set to the following ranges: 3.215-3.231 $\mu$m, 3.252-3.263 $\mu$m, 3.289-3.295 $\mu$m, 3.306-3.311 $\mu$m, 3.610-3.626 $\mu$m, 3.686-3.693 $\mu$m, 3.711-3.727 $\mu$m, and 3.759-3.795 $\mu$m. We calculated a fifth-order polynomial to these regions and took the local optical depths with respect to this continuum. For $^{13}$CO, CH$_3$OH, OCN$^-$, and OCS we scaled laboratory data of simple ice mixtures to match the feature profiles. The profile of the best fitting scaled laboratory mixture and band strengths were used to determine local column densities, as described below. \\

\noindent \textbf{3.4~-~3.6 $\mu$m blended absorption (CH$_3$OH and NH$_3$ $\cdot$ H$_2$O): }
The absorption feature between $\sim$3.35~-~3.6 $\mu$m  is likely caused by a combination of different ices and grain properties.  However there is a distinct peak at 3.53~$\mu$m for both sources indicating the presence of CH$_3$OH ice (C-H stretching mode).\cite{boogert2015observations}  In order to constrain the CH$_3$OH ice abundance, we used the fifth order polynomial local baseline described above to obtain the optical depths for this feature. Previous studies have fit a simple Gaussian to CH$_3$OH along lines of sight toward background stars but this underestimates the red wing in the feature for both lines of sight in this study.\cite{boogert2011ice, Chu2020} We therefore scaled laboratory spectra of pure CH$_3$OH ice at 15K \cite{tvs18_comslab} to  fit the region and minimize the residuals between 3.53 and 3.65~$\mu$m. We did not fit the laboratory data to shorter wavelengths because the use of a local baseline instead of a global baseline cuts off some of the CH$_3$OH ice profile.  Additional absorbing species and  scattering signatures may contribute to this absorption feature. The column densities are calculated by integrating the scaled laboratory optical depths using a band strength $A_{\mathrm{CH_3OH}}=1.6~\times~10^{-16}$~cm~molec$^{-1}$ over the 2.778-3.704~$\mu$m regime\cite{hudgins1993mid} and represent upper limits to the amount of methanol present, due to the potential for additional absorption described above. These results are presented in Table~\ref{tab2} as upper limits, and the fits for both sources are shown in Extended Data Figure 9.
 
The peak near $\sim$3.47~$\mu$m has been previously attributed to the {NH$_3$ $\cdot$ H$_2$O} hydrates but this is still up for debate. \cite{dartois2001search,shimonishi2016vlt}  Nonetheless, we model the feature using a simple Gaussian at this time with a FWHM of 0.1~$\mu$m and a central peak at 3.47~$\mu$m to understand how much this overlapping feature may reduce the column density of the CH$_3$OH.  We model the Gaussian and lab data simultaneously and minimize the residuals of the sum of both fits between 3.4~-~3.65~$\mu$m (Extended Data Figure 9).  When doing this we find that the column densities for CH$_3$OH are lower by $\sim20-30$\% (N$=4.1\times$10$^{17}$~cm$^{-2}$ and N$=4.5\times$10$^{17}$~cm$^{-2}$ for the A$_V=60$ and A$_V=95$ sources, respectively). These column densities agree with those found by using the \texttt{ENIIGMA} fits to the globally determined optical depths.  Further follow-up studies will model the full 3.4~-~3.6~$\mu$m absorption feature and constrain the column densities not only for CH$_3$OH, but also the other possible absorbing species. \\

\noindent \textbf{$^{13}$CO$_2$:} Extended Data Figure 4 shows the observed $^{13}$CO$_2$ feature, around 4.39 $\mu$m, compared with laboratory spectra of CO$_2$ in different ice mixtures, which peak at slightly different wavelengths depending on the ice mixture. The peak of each laboratory spectrum is scaled to the observed $^{13}$CO$_2$ feature at the wavelength corresponding to the peak of the laboratory data. Overall, the band of $^{13}$CO$_2$ in H$_2$O-rich ice reproduces the peak and width of the observed feature. A weak blue shoulder around 4.384 $\mu$m is also noticeable and could possibly be due to a fraction of $^{13}$CO$_2$ mixed in CO. A detailed study of the components that contribute to the 4.39 $\mu$m feature, combined with an analysis of the CO$_2$ bands, can provide more insights about the formation and chemical environment of solid CO$_2$ and its $^{13}$C isotopologue, and it will be the focus of a future work. In this work, we provide an estimate of the $^{13}$CO$_2$ column density assuming that the 4.39 $\mu$m band can be modeled using the laboratory spectrum of a CO$_2$:H$_2$O(1:10) ice at 10 K. The column density of $^{13}$CO$_2$ is derived by scaling the laboratory spectrum to the optical depth of the 4.39 $\mu$m feature. A band strength of  A = $7.8 \times 10^{-17}$ cm molec$^{-1}$ \cite{gerakines1995} is assumed for  $^{13}$CO$_2$ asymmetric stretching. The laboratory data used for the comparison are taken from \cite{Ehrenfreund1997,Ehrenfreund1999}. In these laboratory ices, $^{13}$CO$_2$ is set at ratios of $^{12}$CO$_2$/$^{13}$CO$_2$ $\sim$ 90, which need not be the same ratio in the astronomical targets. To calculate this ratio from the astronomical data, we divide the column densities derived with ENIIGMA for the $^{12}$CO$_2$ feature by those derived here for the $^{13}$CO$_2$ feature, yielding a ratio of $^{12}$CO$_2$/$^{13}$CO$_2$ $\sim$ 69-87 ratio for these two targets.\\

\noindent \textbf{OCN$^-$: } In our analysis of the XCN band a single component fit is used and plotted in Extended Data Figure 7. This is a Gaussian function with peak center at 2165.9 cm$^{-1}$ and FWHM=23 cm$^{-1}$ previously used to reproduce laboratory spectra of OCN$^{-}$ \cite{vanBroekhuizen2004} and the XCN band of embedded young stellar objects.\cite{pendleton1999interstellar,vanBroekhuizen2005,Noble2013a,Noble2017} Only data points on the blue wing of the XCN band are considered to avoid any contributions from the CO-ice band to the fit. The Gaussian profile reproduces the red wing and the component of the XCN band of both targets. In addition, the residuals are negligible, justifying the use of a single component. OCN$^{-}$ ice column densities were estimated by integrating over the fitted Gaussian function and scaling with a band strength $A_{\mathrm{OCN}^-}$ of $1.3 \times 10{^{-16}}$ cm molec$^{-1}$\cite{vanBroekhuizen2004}. The resulting column densities are listed in Table~\ref{tab2},and they are in good agreement with values obtained for quiescent lines of sight in nearby clouds.\cite{vanBroekhuizen2005} \\

\noindent \textbf{$^{13}$CO: } The region around 4.779 $\mu$m shows a weak feature that can be associated with $^{13}$CO (Extended Data Figure 5). This feature is contaminated by the presence of photospheric absorption lines, which makes the feature difficult to integrate cleanly. For this reason, the laboratory spectrum of pure CO ice at 15 K \cite{broekhuizen2006} was scaled to the astronomical data to derive the maximum abundance of this species in the spectra of both background stars. The band strength value is A = $1.3 \times 10^{-17}$ cm molec$^{-1}$. \cite{gerakines1995} \\

\noindent \textbf{OCS:} The region around 4.90 $\mu$m shows tentative detection that can be associated with the CO stretching vibration of the OCS molecule (Extended Data Figure 8). A comparison with laboratory infrared spectra of OCS-containing ices shows that this band in pure OCS ice is too broad compared to the feature seen toward the background stars. Previous studies showed that this absorption band is better modeled by CH$_3$OH:OCS-containing ices \cite{palumbo1995,palumbo1997}. The column densities for OCS were derived using the profile of the OCS mixed in H$_2$O and a band strength value of  A = $1.18 \times 10^{-16}$ cm molec$^{-1}$ \cite{yarnall2022}. \\

\noindent \textbf{Non-detections and upper-limits: } With these high S/N data, we have placed strong constraints on several ice species, including HDO, HCN, CH$_3$CN, H$_2$CO, and H$_2$S. H$_2$CO may still be present at low levels in these spectra, but with the lower resolving power ($R\sim$100) of MIRI LRS FS, it is not possible to separate it from the blue wing of the H$_2$O bending mode at 6 $\mu$m. In contrast, it was clearly detected in a protostar with JWST's MIRI MRS mode (R$\sim$3000) \cite{yang2022corinos}. HDO was tentatively detected with AKARI at $\sim$4.1 $\mu$m with an abundance of 2-10$\%$ relative to H$_2$O towards several protostars and disks \cite{aikawa2012akari}. We do not see an obvious feature there in these spectra, although reliable upper limits can only be obtained after correction for the $^{12}$CO$_2$ blue scattering emission wing. Upper limits for the H$_2$S, CH$_3$CN, and HCN abundances are estimated considering the noise level in the region where the strongest vibrational feature of these molecules absorb. Here, the regions around 3.92 $\mu$m, 4.44 $\mu$m, and 4.76 $\mu$m for H$_2$S, CH$_3$CN, and HCN, respectively. The upper limits are calculated from the root mean squared (RMS) as $\rm N = RMS \times FWHM/A$, where FWHM and A are the full width at half maximum and band strength of the absorption feature in the pure ice, respectively. The resulting 1$\sigma$ upper limits in the abundances w.r.t. H$_2$O ice is less than 1 \% for H$_2$S in NIR38. Data covering this region is not yet available for J110621. For HCN the upper limits w.r.t. H$_2$O ice is less than 1\% for both sources. For CH$_3$CN the value is less than 2\% for both sources. The FWHM and band strengths for the pure ices are taken from \cite{yarnall2022,rachid2022,gerakines2022}.\\

\noindent \textbf{The location of the background stars in their larger scale environments: } Complementary information about the larger scale environment is critical when interpreting the column densities inferred from the ice observations. Extended Data Figure 10 shows a map of the H$_2$ column density maps extracted from the larger scale Chamaeleon maps\cite{alvesdeoliveria14} created based on far-infrared data 70 to 500~$\mu$m from the \textit{Herschel} Space Observatory's Gould Belt survey\cite{andre10}. The maps clearly show the decrease of the column density from the peak near the Class~0 protostar ChamI-MMS with a more extended structure emcompassing also the clump Cha1-C2\cite{belloche2011a}. The $A_V \approx 95$ star at a projected distance of 6600~au is located in the direction of this core, while the $A_V \approx 60$ star at a projected distance of 5600~au is located in a direction orthogonal to this structure from the Class~0 protostar\cite{jin2022ice}. The H$_2$ column densities toward the two background stars are similar within $\approx$10\%, suggesting that the local conditions are similar, despite the difference in $A_V$. The $A_V$ we use was derived from average giant star colors\cite{jin2022ice}; from these JWST spectra, a more detailed fit taking into account the spectral type of these background stars will soon be possible, which may reduce the difference in $A_V$. If the discrepancy remains, the difference in $A_V$ could represent local radial extensions of the cloud along the line of sight or a superposition of additional clouds along the line of sight. Complementary observations, e.g. of gas-phase line tracers, are needed to assess whether there are differences in the densities, and thereby, e.g., the time-scales for freeze-out, toward the two differ significantly. \\

\noindent \textbf{Calculation of the icy C, O, N, and S budgets: }
The column of molecular hydrogen is calculated for each line of sight as $N_{H2}\sim1.0\times10^{21}cm^{-2}\ A_V$ \cite{lacy2017}. Assuming cosmic abundances for the combined volatile and refractory abundances in the interstellar medium (ISM)\cite{przybilla2008}, the molecular hydrogen column can be converted into expected bulk budgets of C, O, N, and S. To determine what fraction of these budgets our ices represent, we summed the column densities of all C-bearing, O-bearing, N-bearing, and S-bearing ice species. For the O-bearing species, we doubled the column densities of $^{12}CO_2$, $^{13}CO_2$, and $SO_2$ to account for the two oxygen atoms. For both NIR38 and J110621, we see only 19\% of the total O-budget, 19\% and 14\%, respectively, of the C-budget, and 13\% of the N-budget and 1\% of the S-budget for both. If we assume the $N_{H2}/N_{CO}$ conversion for molecular clouds\cite{lacy2017}, then the expected amount of total CO towards NIR38 and J110621 are 1.08$\times$10$^{19}$ cm$^{-2}$ and 1.71$\times$10$^{19}$ cm$^{-2}$, respectively.


\section*{Data Availability}
Our raw data are available at the STScI MAST JWST archive, and our enhanced spectra are available as part of our ERS science enabling product deliverables at the following Zenodo DOI: 10.5281/zenodo.7501239

\section*{Code Availability}
The \texttt{ENIIGMA} global fitting tool\cite{Rocha2021} is publicly available on GitHub at the following URL: https://github.com/willastro/ENIIGMA-fitting-tool


\section*{Acknowledgements}
The Ice Age ERS team would like to thank our support team at STScI (William Januszewski, Beth Sargent, Norbert Pirzkal, and Mike Engesser) for their technical suggestions and improvements to the program since 2017.  We would also like to thank the anonymous referees for suggestions that improved the manuscript. 
MKM acknowledges financial support from the Dutch Research Council (NWO; grant VI.Veni.192.241). 
MR acknowleges support from the Netherlands Research School for Astronomy (NOVA).
SI, HL, and EvD acknowledge support from the Danish National Research Foundation through the Center of Excellence “InterCat” (Grant agreement no.: DNRF150).
EvD acknowledges support from ERC grant 101019751 MOLDISK.
The research of LEK is supported by a research grant (19127) from VILLUM FONDEN.
Part of this research was carried out at the Jet Propulsion Laboratory, California Institute of Technology, under a contract with the National Aeronautics and Space Administration (DL).
FS acknowledges funding from JWST/NIRCam contract to the University of Arizona, NAS5-02105.
ACAB acknowledges support from the Space Telescope Science Institute for program JWST-ERS-01309.019.
JE acknowledges support from the Space Telescope Science Institute for program JWST-ERS-01309.019.
LEUC's research was supported by an appointment to the NASA Postdoctoral Program at the NASA Ames Research Center, administered by Oak Ridge Associated Universities under contract with NASA.
D. H. is supported by Center for Informatics and Computation in Astronomy (CICA) grant and grant number 110J0353I9 from the Ministry of Education of Taiwan.  DH acknowledges support from the National Technology and Science Council of Taiwan through grant number 111B3005191. 
M.N.D. acknowledges the Swiss National Science Foundation (SNSF) Ambizione grant no. 180079, the Center for Space and Habitability (CSH) Fellowship, and the IAU Gruber Foundation Fellowship.
I.J.-S. acknowledges financial support from grant No. PID2019-105552RB-C41 by the Spanish Ministry of Science and Innovation/State Agency of Research MCIN/AEI/10.13039/501100011033.
This work was supported by a grant from the Simons Foundation (686302, KIÖ) and an award from the Simons Foundation (321183FY19, KIÖ).
JKJ acknowledges support from the Independent Research Fund Denmark (grant number 0135-00123B).
ZLS acknowledges financial support from the Royal Astronomical Society through the E.A Milne Travelling Fellowship.
JAN and ED acknowledge support from French Programme National “Physique et Chimie du Milieu Interstellaire” (PCMI) of the CNRS/INSU with the INC/INP, co-funded by the CEA and the CNES.

\section*{Author contributions} 
MKM originated the proposal, designed the observations, co-managed the team, determined the feature optical depths and wrote much of the main text. WR performed global and local fitting to determine the column densities, including the error analysis, and wrote part of the Methods section. KP contributed to the observational design, reduced and optimized the NIRSpec data, wrote part of the Methods section, and commented on the draft. NC reduced and optimized the MIRI LRS data to allow for the global fitting and wrote part of the Methods section. LEUC performed the local fitting of the methanol + hydrates band, wrote part of the Methods section, and commented on the draft. ED wrote part of the discussion and made suggestions for the analysis. TL wrote portions of the results section and reorganized the draft. JAN contributed to the original proposal, wrote portions of results section, and made suggestions for the analysis. YJP managed the Overleaf file, wrote part of the results section, and made suggestions for the local fitting. GP locally fit the OCN$^-$ feature, wrote part of the Methods section, and commented on the draft. DQ managed the Overleaf file and suggested parts of the results and discussion sections. MGR did the local fitting of the $^{13}$CO$_2$, $^{13}$CO and OCS features, determined the upper limits, and wrote part of the Methods section. ZLS and FS reduced the NIRCam data, with contributions to the reduction scripts from HD, and wrote part of the Methods section. TB benchmarked the NIRSpec spectra to validate them. ACAB helped to design the original program, co-managed the team, organized the NIRCam analysis, and commented on the draft. WAB, PC, SBC, HC, MND, EE, JE, HF, RTG, DH, SI, IJS, MJ, JKJ, LEK, DCL, MRSM, BAM, GJM, KIO, MEP, TS, JAS, EFvD, and HL commented on the draft. ZS, FS, EE, JE, HF, and TS also contributed to the observational design and analysis of the NIRCam data.  HL helped motivate the original proposal, co-managed the team, and organized the laboratory data used for the analysis. All authors participated in discussion of the observations, analysis and interpretation of the results.

\section*{Competing Interests}
The authors declare no competing financial interests.\\ 
\clearpage


\begin{table}[h]
\begin{center}
\begin{minipage}{600pt}
\caption{Absorption features of molecules in ices and dust features observed towards NIR38 (A$_V\sim$60) and J110621 (A$_V\sim$95).}
\label{tab1}%
\begin{tabular}{@{}llllll@{}}
\toprule
$\lambda$ ($\mu$m) & $\nu$ (cm$^{-1}$) &  Species  & Identification$^a$ & \multicolumn{2}{c}{Detection}\\ 
        &  &        & &    NIR 38  & J110621  \\
\midrule
2.69 & 3708 & CO$_2$ & combination  & \cmark & \bf{...}\\
2.73 & 3664 & H$_2$O & O$-$H dangling bond  & \cmark & \bf{...} \\
3.0 & 3330 & H$_2$O & O$-$H stretch  & \cmark & \cmark \\
3.24 & 3249 & CH$_3$OH & O$-$H stretch  & \cmark & \cmark \\
3.32 & 3012 & CH$_4$ & C$-$H stretch & \cmark & \cmark \\
3.47 & 2881 & Ammonia hydrate & NH$_3$ $\cdot$ H$_2$O & \bf{!} & \bf{!}\\
3.32$-$3.64 & 3012$-$2890 & CH$_3$OH & C$-$H asym. str. + overt. & \cmark & \cmark\\
3.92 & 2548 & H$_2$S & S$-$H & \xmark & \bf{...} \\
4.07 & 2457 & HDO & O$-$D str. & \xmark & \xmark\\
4.17$-$4.77 & 2400$-$2100 & H$_2$O & combination & \cmark & \cmark\\
4.27 & 2340 & $^{12}$CO$_2$ & C$-$O str. & \cmark & \cmark\\
4.38 & 2280 & $^{13}$CO$_2$ & C$-$O str. & \cmark & \cmark\\ 
4.44 & 2252 & CH$_3$CN & C$-$N str. & \xmark & \xmark\\
4.59 & 2175 & OCN$^{-}$ & C$-$N str. & \cmark & \cmark\\
4.67 & 2140 & $^{12}$CO & C$-$O str. & \cmark & \cmark\\
4.76 & 2100 & HCN & C$\equiv$N str. & \xmark & \xmark\\
4.78 & 2090 & $^{13}$CO & C$-$O str. & \cmark & \cmark\\
4.90 & 2040 & OCS & C$-$O str. & \cmark & \cmark\\
6.0 & 1666 & H$_2$O & bending & \cmark & \cmark\\
6.85 & 1459 & CH$_3$OH & CH$_3$ def. & \cmark & \cmark\\
6.85 & 1459 & NH$_4^+$ & N$-$H str. & \cmark & \cmark\\
6.9-7.5 & 1449-1333 & Unidentified absorption & COMs functional groups? & \cmark & \cmark \\
7.24 & 1384 & CH$_3$CH$_2$OH? & CH$_3$ def. & \bf{!} & \bf{!}\\
7.43 & 1362 & CH$_3$CHO? & CH$_3$ def. + CH wag. & \bf{!} & \bf{!}\\
7.60 & 1318 & SO$_2$ & S$-$O str. & \bf{!} & \bf{!}\\
7.71 & 1300 & CH$_4$ & C$-$H str. & \cmark & \cmark\\
8.86 & 1131 & CH$_3$OH & CH$_3$ rock & \cmark & \cmark\\
9.01 & 1110 & NH$_3$ & umbrella & \cmark & \cmark\\
9.74 & 1025 & CH$_3$OH & C$-$O str. & \cmark & \cmark\\
9.80 & 1020 & Silicate & Si$-$O str. & \cmark & \cmark\\
11.0 & 910 & H$_2$O & libration wing & \cmark & \cmark\\
\bottomrule
\end{tabular}
\footnotetext{$^a$Symbol legend: \cmark - observed, \xmark - not observed, {\bf{!}} - possibly observed, {\bf{...}} - insufficient data}
\end{minipage}
\end{center}
\end{table}
\clearpage

\begin{table}[h]
\begin{minipage}{600pt}
\caption{Integrated optical depths and column densities of molecules in ices observed towards AV60 and AV95 sources.}
\label{tab2}%
\renewcommand{\arraystretch}{1.3}
\begin{tabular}{@{}lllllll@{}}
\toprule
Species  & $\nu$ (cm$^{-1}$) & $A$ (cm molec$^{-1}$)&\multicolumn{2}{c}{$\int\tau_{\nu} d\nu$ $^a$} & \multicolumn{2}{c}{$N_{\rm{ice}}$ [$\times$10$^{18}$ cm$^{-2}$]$^b$}\\
\cline{4-7}
                    & & &    $\rm{A_V}$=60 & $\rm{A_V}$=95        & $\rm{A_V}$=60 & $\rm{A_V}$=95 \\
\midrule
H$_2$O   & 3330  & 2 $\times$ 10$^{-16}$ [ref.\cite{gerakines1995}] & 1376.07  & 2676.12 &  6.88$_{3.70}^{12.5}$ (6.93)  & 13.38$_{7.83}^{17.27}$ (13.17)\\
$^{12}$CO & 2140 & 1.1 $\times$ 10$^{-17}$ [ref.\cite{gerakines1995}] & 32.56 &  40.48  & 2.96$_{1.86}^{4.66}$ (3.22) & 3.68$_{2.48}^{5.46}$ (3.94)\\
$^{13}$CO & 2090 & 1.0 $\times$ 10$^{-17}$ [ref.\cite{gerakines1995}] & 0.32 &  0.31  & 0.03$_{0.02}^{0.04}$ (0.02) & 0.02$_{0.01}^{0.03}$ (0.02)\\
$^{12}$CO$_2$ & 2340 & 1.1 $\times$ 10$^{-16}$ [ref.\cite{gerakines1995}] & 151.8 &  191.4  & 1.38$_{0.77}^{1.97}$ (1.36) & 1.74$_{1.09}^{2.36}$ (1.62)\\
$^{13}$CO$_2$ & 2280 & 7.1 $\times$ 10$^{-17}$ [ref.\cite{gerakines1995}] &  1.42 &  2.30  & 0.02$_{0.01}^{0.03}$ (0.03) & 0.02$_{0.02}^{0.04}$ (0.03)\\
CH$_3$OH$^c$ & 2830 & 1.3 $\times$ 10$^{-16}$ [ref.\cite{hudgins1993mid}] & 53.3 \{68.9\} &  58.5 \{84.5\}  & 0.41 \{0.53\} & 0.45 \{0.65\}\\
CH$_3$OH & 1025 & 1.8 $\times$ 10$^{-17}$ [ref. \cite{bouilloud2015}] & 10.98 &  9.18  & 0.61$_{0.28}^{0.95}$ (0.54) & 0.51$_{0.24}^{1.08}$ (0.49) \\
NH$_3$ & 1110 & 2.1 $\times$ 10$^{-17}$ [ref.\cite{bouilloud2015}] & 6.31 &  13.86  & 0.30$_{0.21}^{0.97}$ (0.41) & 0.66$_{0.48}^{1.11}$ (0.68)\\
CH$_4$ & 1303 & 8.4 $\times$ 10$^{-18}$ [ref.\cite{bouilloud2015}] & 1.51 &  2.11  & 0.18$_{0.14}^{0.23}$ (0.16) & 0.25$_{0.16}^{0.28}$ (0.28)\\
OCN$^-$ & 2175 & 1.3 $\times$ 10$^{-16}$ [ref.\cite{vanBroekhuizen2005}] & 2.58$^d$ &  4.11$^d$  & 0.02 & 0.03\\
NH$_4^+$ & 1459 & 4.4 $\times$ 10$^{-17}$ [ref.\cite{Schutte2003}] & 25.08$^d$ &  34.32$^d$  & 0.57 & 0.78\\
OCS & 2040 & 1.2 $\times$ 10$^{-16}$ [ref.\cite{yarnall2022}] & 1.18$^d$ &  2.36$^d$  & 0.01 & 0.02\\
SO$_2$ & 1310 & 3.4 $\times$ 10$^{-17}$ [ref.\cite{boogert1997}] & 0.11$^d$ &  0.16$^d$  & 0.0034 & 0.0047\\
\hline
 & & & \multicolumn{4}{c}{1$\sigma$ upper limits}\\
\cline{4-7}
H$_2$S & 2548 & 1.7 $\times$ 10$^{-17}$[ref.\cite{yarnall2022}] & 0.64$^e$ &  ...  & 0.04 & ...\\
HCN & 2100 & 1.0 $\times$ 10$^{-17}$[ref.\cite{gerakines2022}] & 0.66$^e$ &  0.88$^e$  & 0.06 & 0.09\\
CH$_3$CN & 2252 & 1.9 $\times$ 10$^{-18}$[ref.\cite{rachid2022}] & 0.26$^e$ &  0.35$^e$  & 0.14 & 0.19\\
\bottomrule
\end{tabular}
\footnotetext{$^a$When not indicated, these values are based on the global fit.}
\footnotetext{$^b$ Upper and lower values are from 3$\sigma$ confidence intervals. Values inside the parenthesis are calculated from local fits.\\ See Supplementary Information.}
\footnotetext{$^c$ Calculations performed on optical depth data after local continuum extraction around 3.5~$\mu$m. See Extended Data Figure 9. Values inside \\the curly brackets are obtained excluding the ammonia hydrate effect.}
\footnotetext{$^d$ Values from the local fits (see Supplementary Information).}
\footnotetext{$^e$ $\int \tau_{\nu}$d$\nu$ = RMS $\times$ FWHM.}
\end{minipage}
\end{table}
\clearpage

\begin{table}[h]
\caption{\label{tab3} Laboratory data tested in the global fit performed with \texttt{ENIIGMA}.}
\renewcommand{\arraystretch}{1.0}
\centering 
\begin{tabular}{lcc}
\hline\hline
Label/Temp. & Temperature (K) & Database/Reference\\
\hline
H$_2$O & 15~K &  LIDA\cite{Oberg2007}\\
NH$_3$ & 10~K &  LIDA\cite{taban2003}\\
CH$_4$ & 10~K &  LIDA\cite{rachid2020}\\
CO & 12~K & LIDA\cite{vanBroekhuizen2004}\\
CO$_2$ & 12~K & LIDA\cite{vanBroekhuizen2004}\\
CH$_3$OH & 10~K & LIDA\cite{tvs18_comslab}\\
NH$_3$:CH$_3$OH (1:1) & 12~K & UNIVAP\cite{Rocha2021}\\
H$_2$O:NH$_3$ (10:1.6) & 10~K & ...\\
H$_2$O:CO$_2$ (10:1) & 10~K & LIDA\cite{Ehrenfreund1997}\\
H$_2$O:CO$_2$ (1:10) & 10~K & LIDA\cite{Ehrenfreund1997}\\
H$_2$O:CO$_2$ (1:6) & 10~K & LIDA\cite{Ehrenfreund1997}\\
H$_2$O:CO$_2$ (1:1) & 10~K & LIDA\cite{Ehrenfreund1997}\\
H$_2$O:CO (20:1) & 16~K & NASA/Ames\cite{Hudson1999}\\
H$_2$O:CH$_4$ (20:1) & 15~K & NASA/Ames\cite{Hudson1999}\\
H$_2$O:CO$_2$:CH$_4$ (10:1:1) & 12~K & UNIVAP\cite{Rocha2017}\\
H$_2$O:CH$_3$OH:CO$_2$:CH$_4$ (0.6:0.7:1:0.1) & 10~K & LIDA\cite{Ehrenfreund1999}\\
H$_2$O:CH$_3$OH:CO$_2$ (9:1:2) & 10~K & LIDA\cite{Ehrenfreund1999}\\
H$_2$O:CH$_3$OH:CO:NH$_3$ (100:50:1:1) & 10~K & NASA/Ames\cite{hudgins1993mid}\\
H$_2$O:CH$_3$OH (10:0.8) & 10~K & ...\\
CO$_2$:CH$_3$OH (1:1) & 10~K & LIDA\cite{Ehrenfreund1997}\\
CO:CO$_2$ (1:1) & 15~K & LIDA\cite{vanBroekhuizen2004}\\
CO:CH$_3$OH (4:1) & 15~K & LIDA\cite{Cuppen2011}\\
\hline
\end{tabular}
\end{table}

\clearpage


\clearpage

\clearpage
\appendix

\section{Extended Data}
\begin{figure}[htb!]
\centering
\includegraphics[width=\hsize]{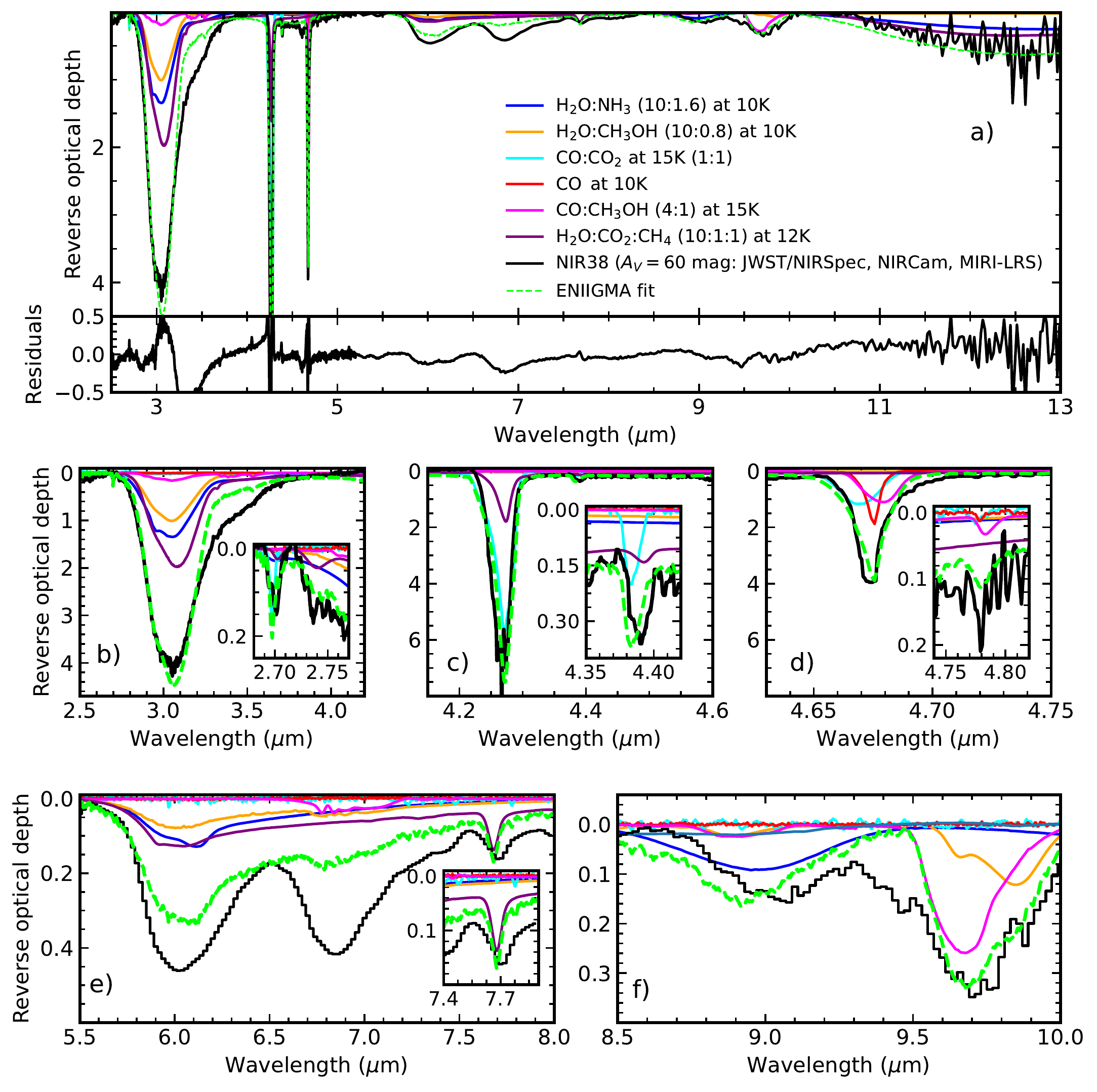}
\caption*{Extended Data Figure 1: \textbf{Global fit of the combined spectrum for NIR38.} Combined NIRSpec and MIRI/LRS spectrum of the NIR38 source (black), with the \texttt{ENIIGMA} fitting tool model (green). Each component in the fit is colour-coded. Panel {\it a} shows the entire range between 2.5 and 13~$\mu$m and the residuals of the fit. Panels {\it b}-{\it f} show a zoom-in of selected ranges corresponding to the major ice components. Small insets show the fit of $^{12}$CO$_2$ (Panel {\it b}), $^{13}$CO$_2$ (Panel {\it c}), $^{13}$CO (panel {\it d}) and CH$_4$ (panel {\it e}).}
\label{ed_fig1}
\end{figure}

\begin{figure}[htb!]
\centering
\includegraphics[width=\hsize]{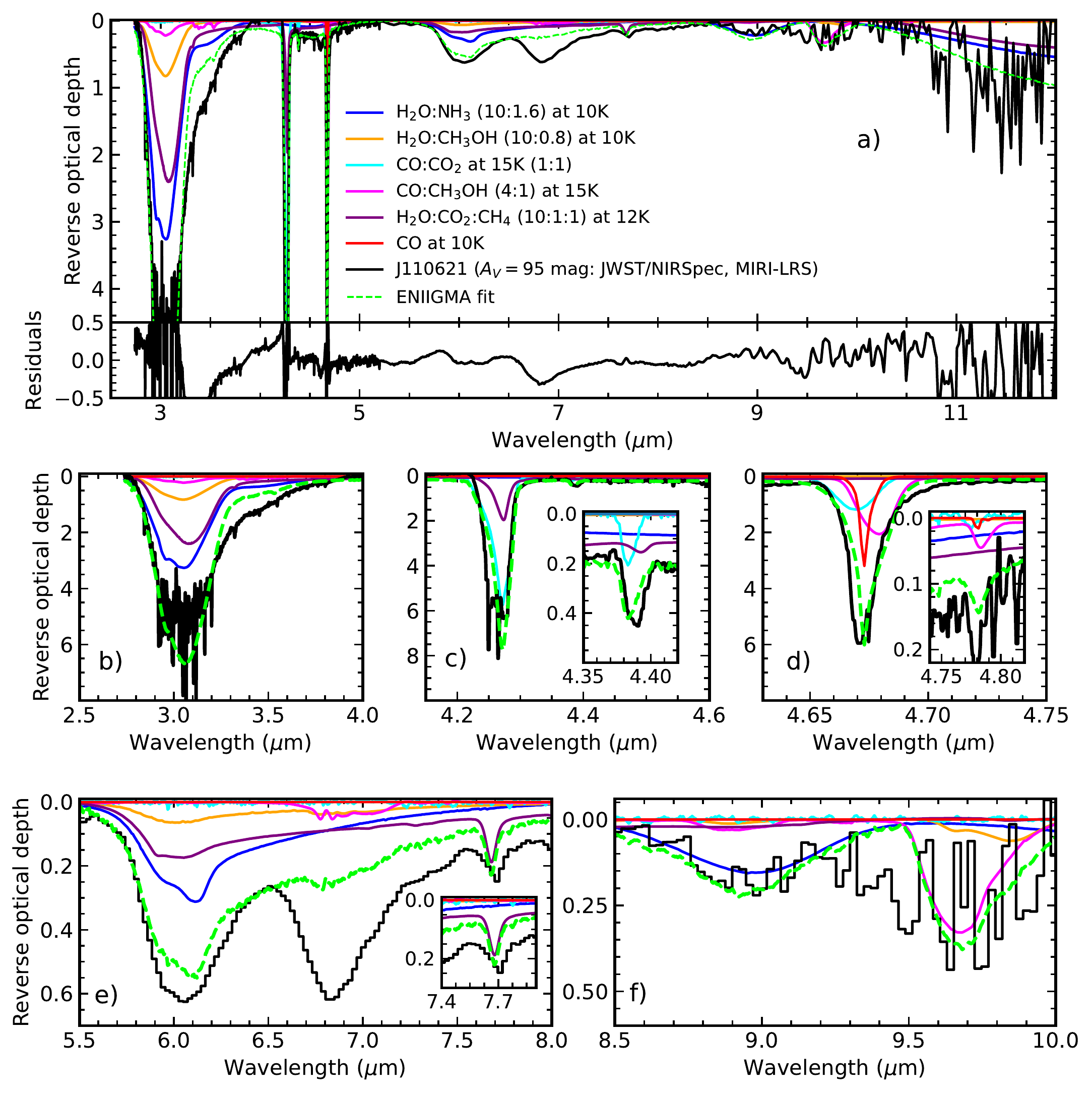}
\caption*{Extended Data Figure 2: \textbf{Global fit of the combined spectrum for J110621.} Combined NIRSpec and MIRI/LRS spectrum of the J110621 source (black), with the \texttt{ENIIGMA} fitting tool model (green). Each component in the fit is colour-coded. Panel {\it a} shows the entire range between 2.5 and 13~$\mu$m and the residuals of the fit. Panels {\it b}-{\it f} show a zoom-in of selected ranges corresponding to the major ice components. Small insets show the fit of $^{13}$CO$_2$ (Panel {\it c}), $^{13}$CO (panel {\it d}) and CH$_4$ (panel {\it e}).}
\label{ed_fig2}
\end{figure}

\begin{figure}[htb!]
\centering
\includegraphics[width=\hsize]{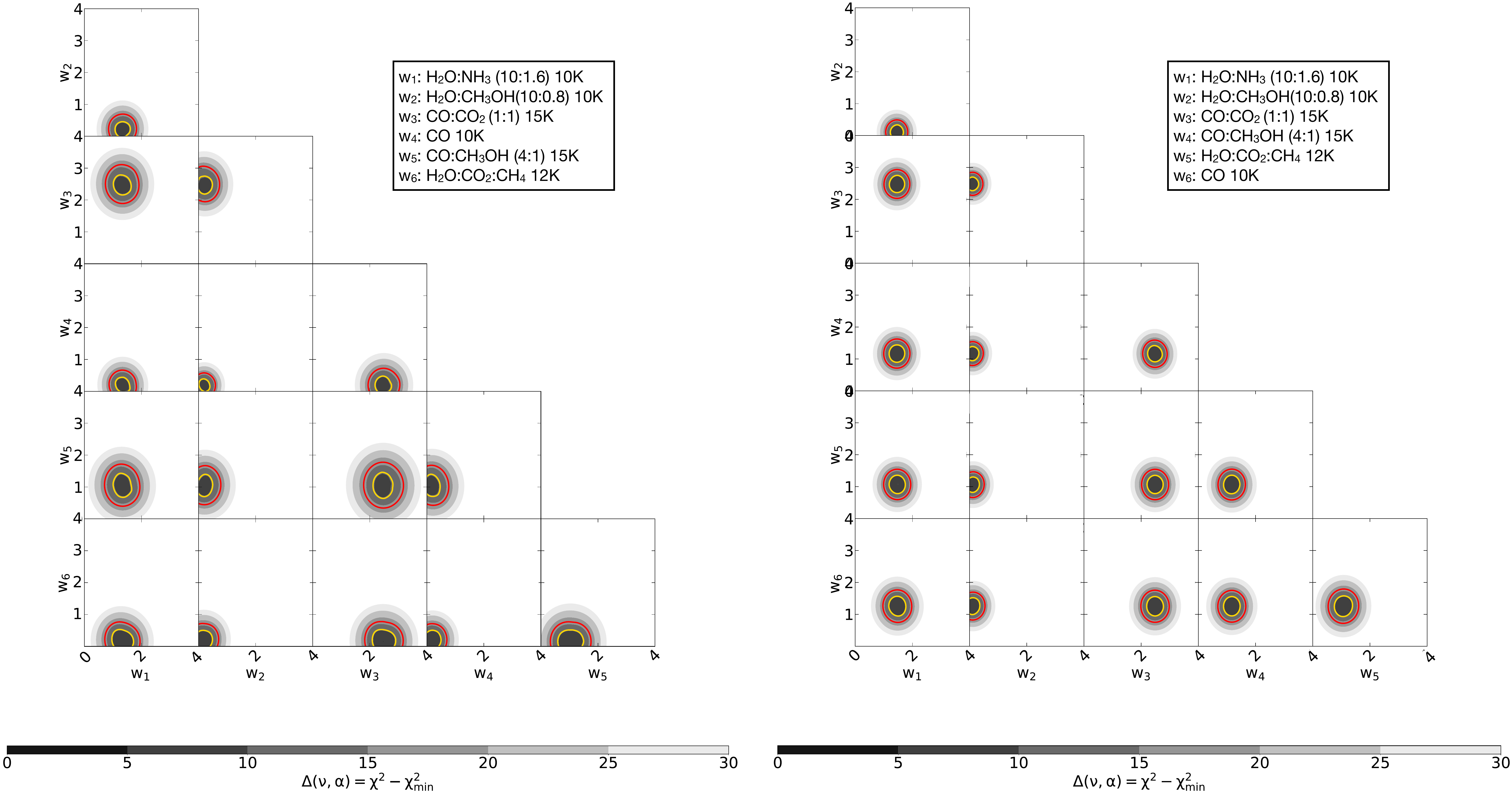}
\caption*{Extended Data Figure 3: \textbf{Confidence interval analysis for the global fits to NIR38 and J110621.} Corner plot showing the confidence interval analysis of the coefficients in the linear combination. The grey-scale contours show the differences in the $\chi^2$ maps ($\Delta$) which depends on the degree of freedom ($\nu$) and the statistical significance ($\alpha$). The yellow and red line contours indicate 2 and 3$\sigma$ confidence intervals. The {\it left} and {\it right} plots are for $A_V = 60$ and $A_V = 95$ sources, respectively. Note that the ice species assigned to w1-w6 is automatically determined and differs between the left and right panels.}
\label{ed_fig3}
\end{figure}

\begin{figure}[htb!]
\centering
\includegraphics[totalheight=8cm]{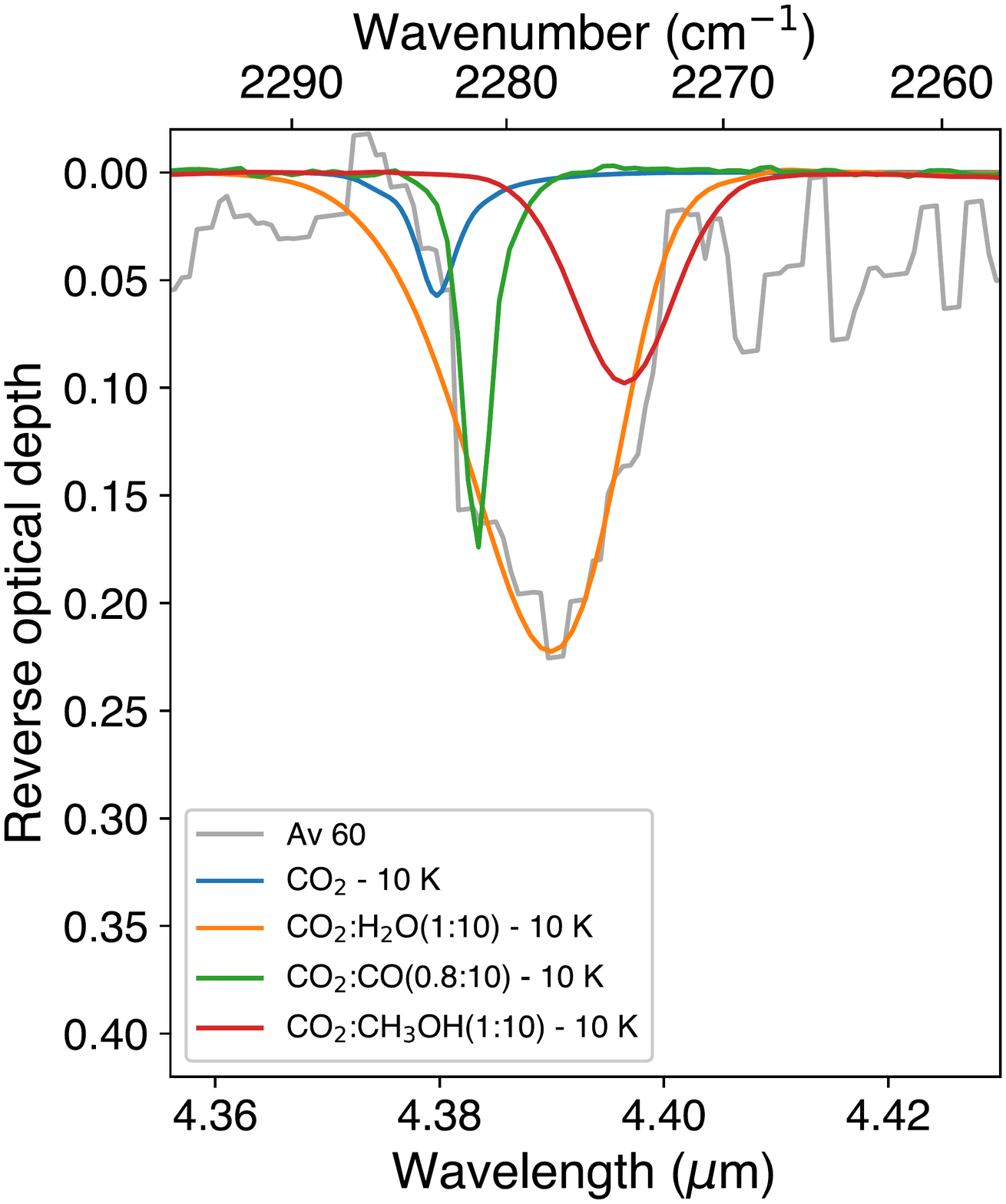}
\includegraphics[totalheight=8cm]{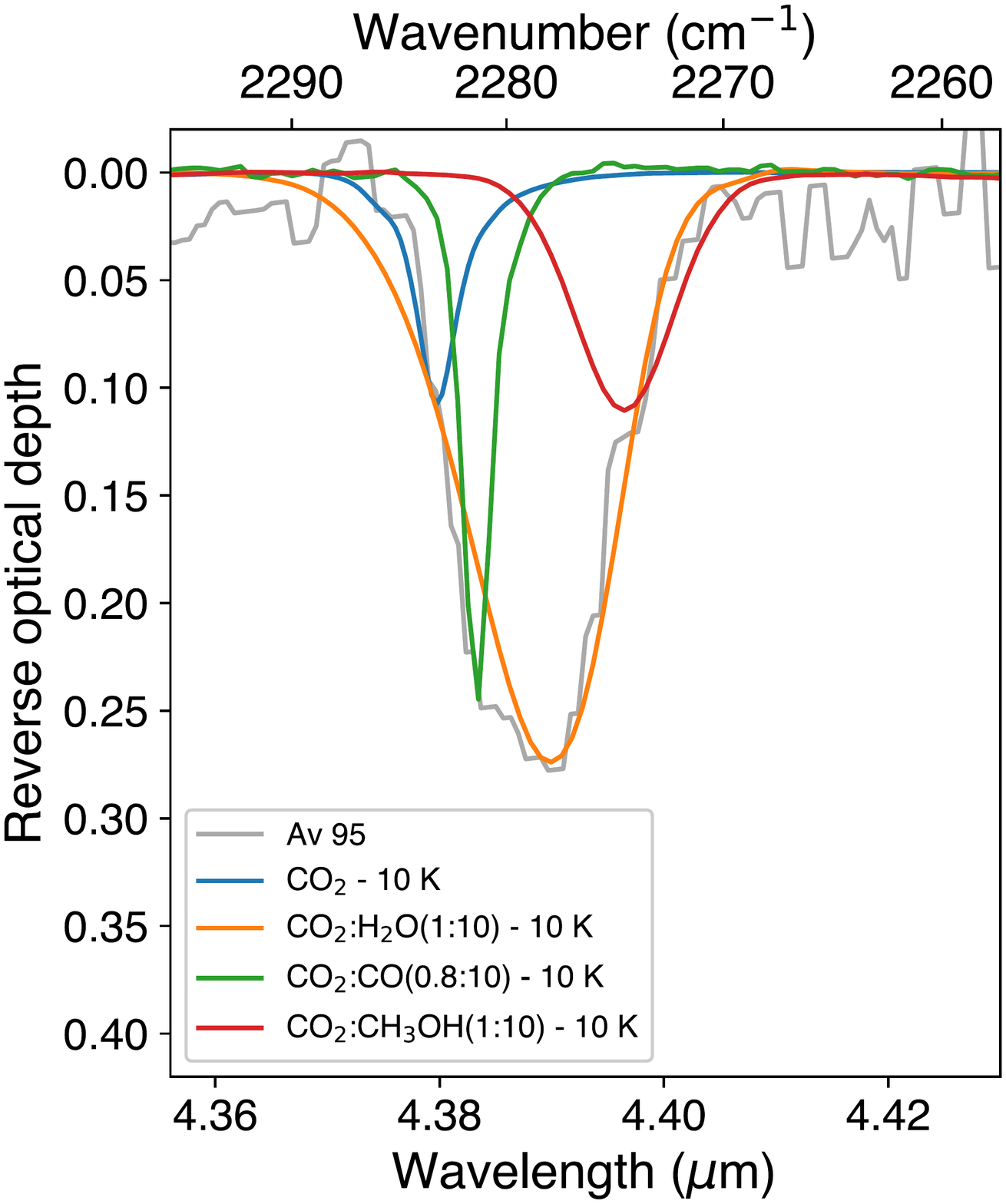}
\caption*{Extended Data Figure 4: Observed absorption profile of the $^{13}$CO$_2$ asymmetric stretching, around 4.39 $\mu$m, in NIR38 (left panel) and J110621 (right panel). To demonstrate the ice chemical environment that best reproduces the observed feature peak, the colored curves show the scaled profiles of $^{13}$CO$_2$ in laboratory spectra of the following ice mixtures at 10 K: pure CO$_2$ (blue), H$_2$O:CO$_2$ (orange), CO$_2$:CO (green), and CO$_2$:CH$_3$OH (red). In all the ice mixtures, CO$_2$ is diluted in a ratio of $\sim$1:10, with $^{12}$CO$_2$/$^{13}$CO$_2$ $\sim$ 90.}
\label{ed_fig4}
\end{figure}

\begin{figure}[htb!]
\centering
\includegraphics[totalheight=8cm]{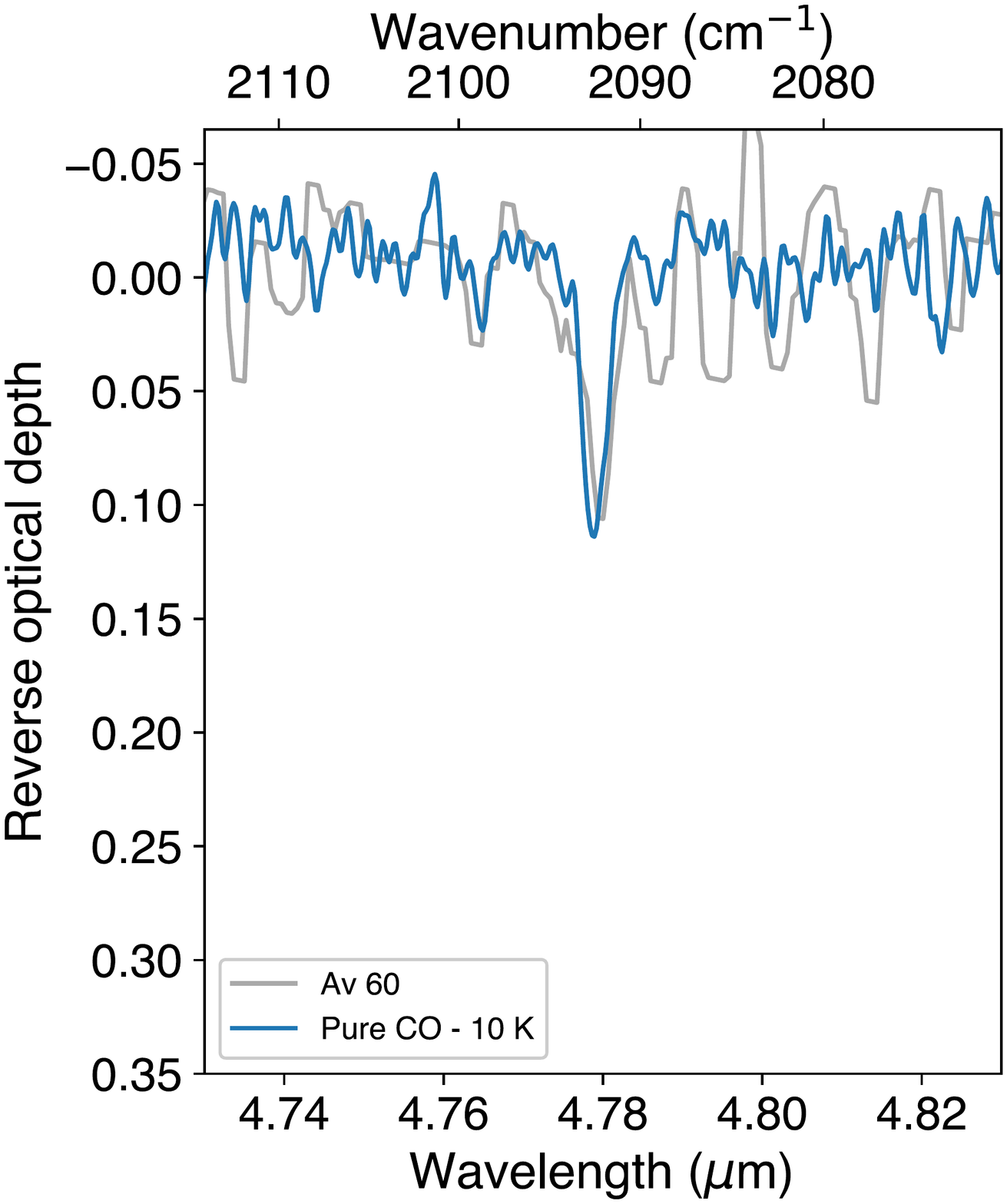}
\includegraphics[totalheight=8cm]{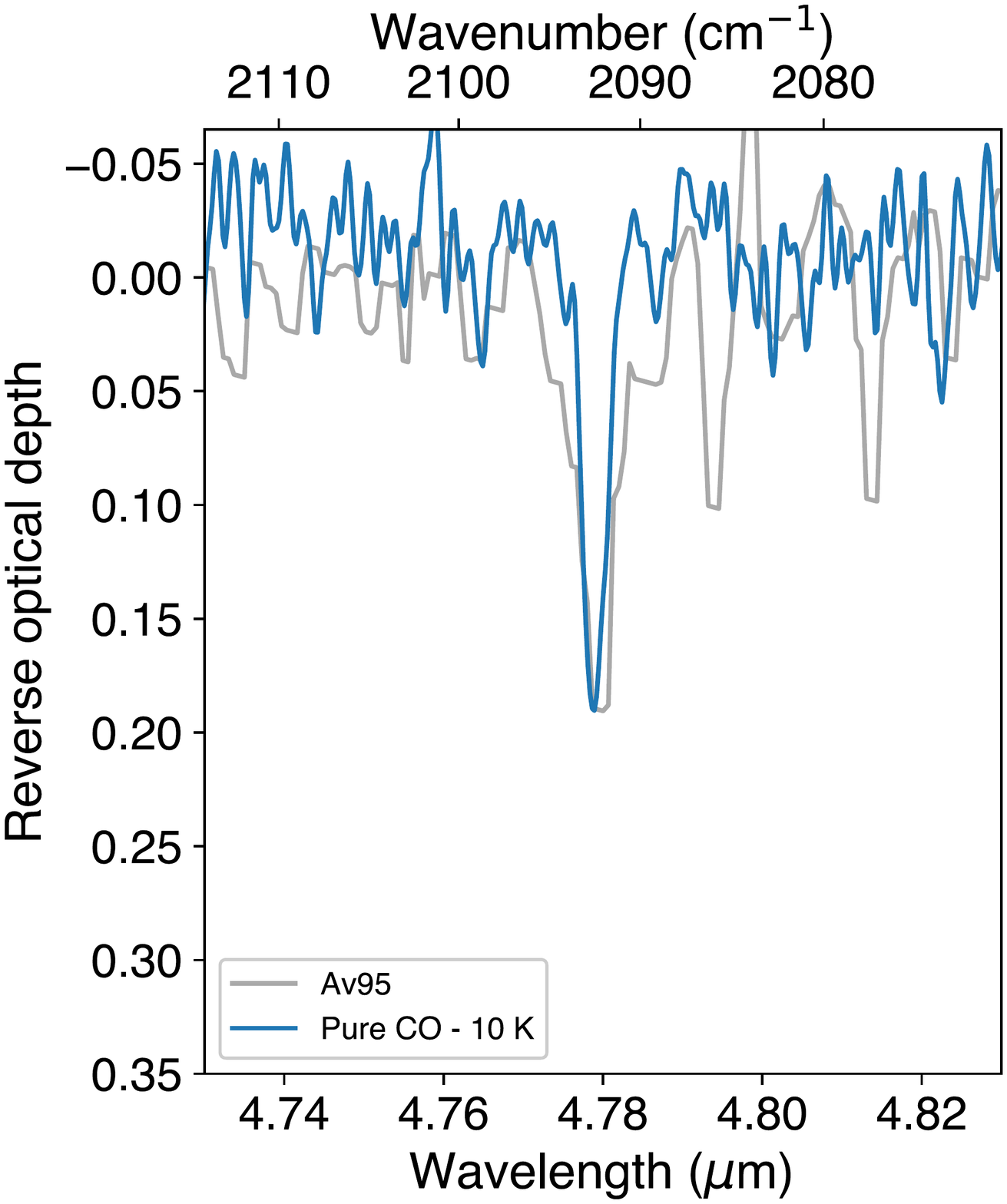}

\caption*{Extended Data Figure 5: Observed absorption profile of the $^{13}$CO stretching, around 4.78 $\mu$m,  in the  Av = 60 (left panel) and Av = 95 (right panel) sources. The laboratory spectra of pure $^{13}$CO ice at 10 K is also shown in blue.}
\label{ed_fig5}
\end{figure}

\begin{figure}[htb!]
\centering
\includegraphics[width=\hsize]{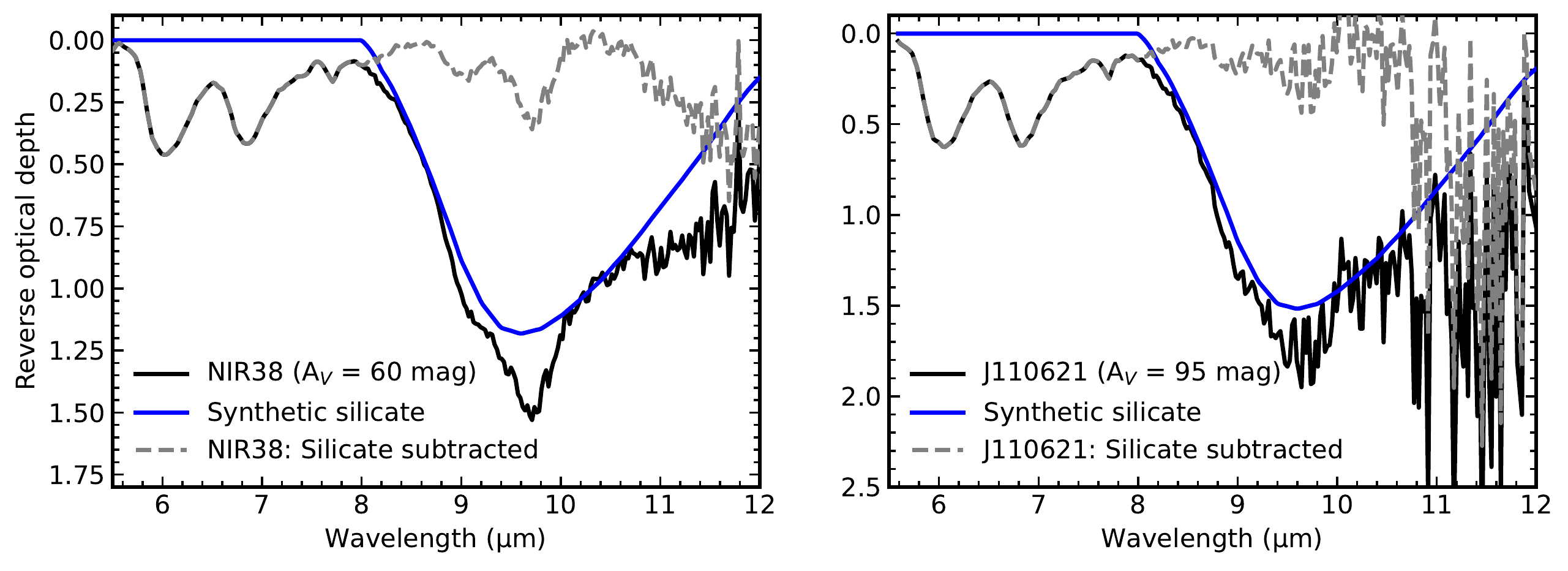}
\caption*{Extended Data Figure 6: \textbf{Silicate subtraction during optical depth calculation for NIR38 and J110621.} MIRI/LRS spectrum of the two background stars before (black) and after (blue) silicate subtraction. The grey dashed line is the synthetic silicate spectrum used to remove the silicate absorption toward the background stars.}
\label{ed_fig6}
\end{figure}

\begin{figure}[htb!]
\centering
\includegraphics[totalheight=7cm]{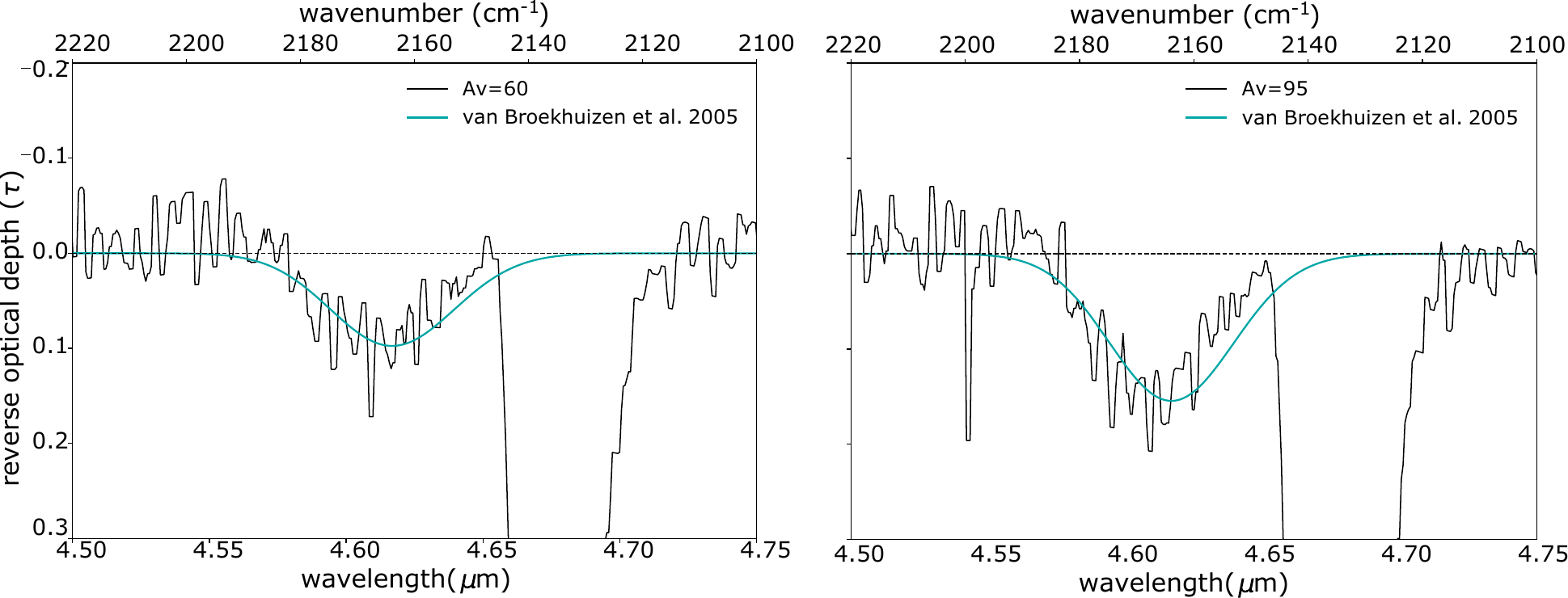}
\caption*{Extended Data Figure 7: Observed absorption profile of the OCN$^-$ feature around 4.62 $\mu$m, in the  Av = 60 (left panel) and Av = 95 (right panel) sources. A gaussian fit using the parameters found in the literature \cite{vanBroekhuizen2004} is also shown.}
\label{ed_fig7}
\end{figure}

\begin{figure}[htb!]
\centering
\includegraphics[totalheight=8cm]{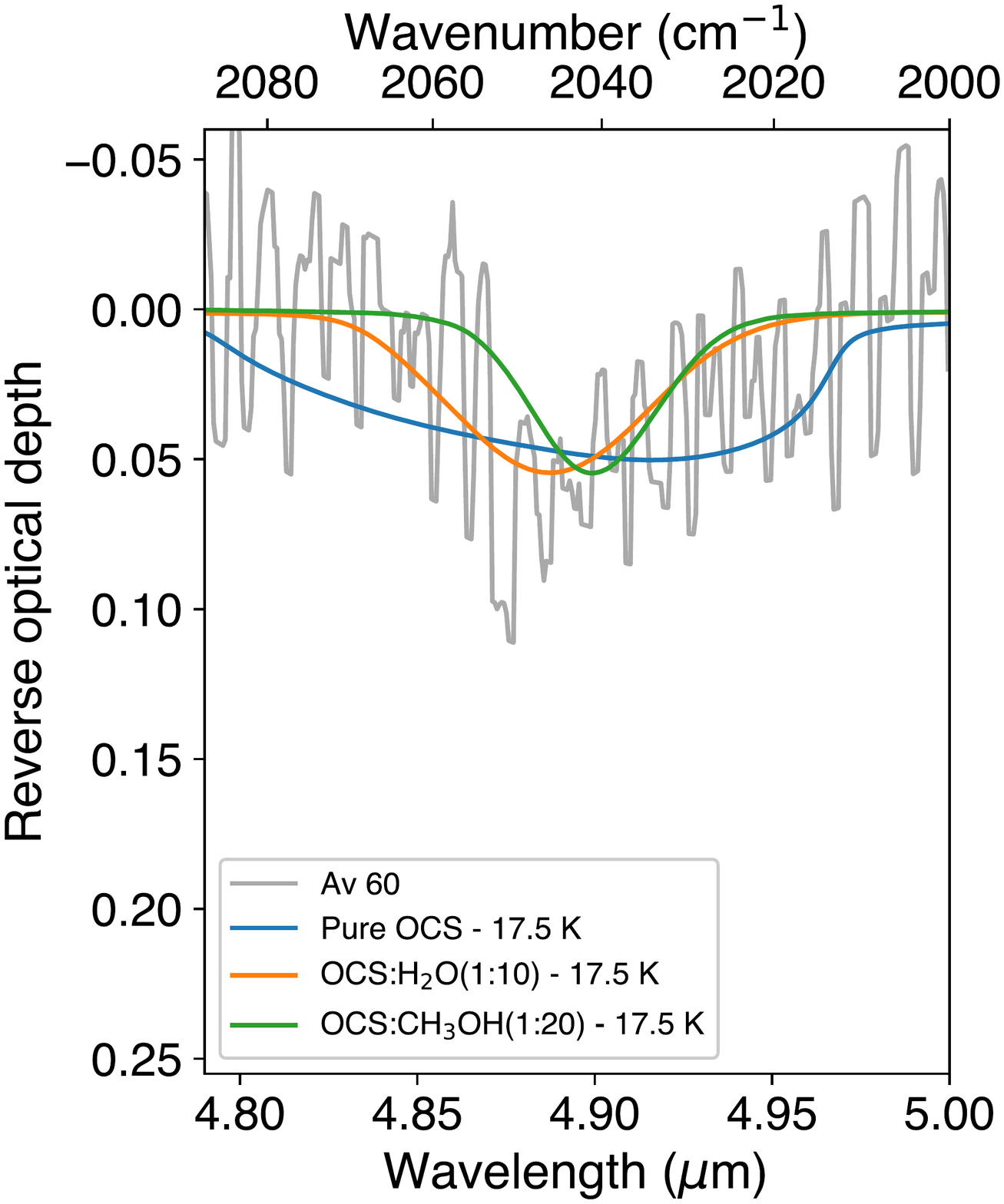}
\includegraphics[totalheight=8cm]{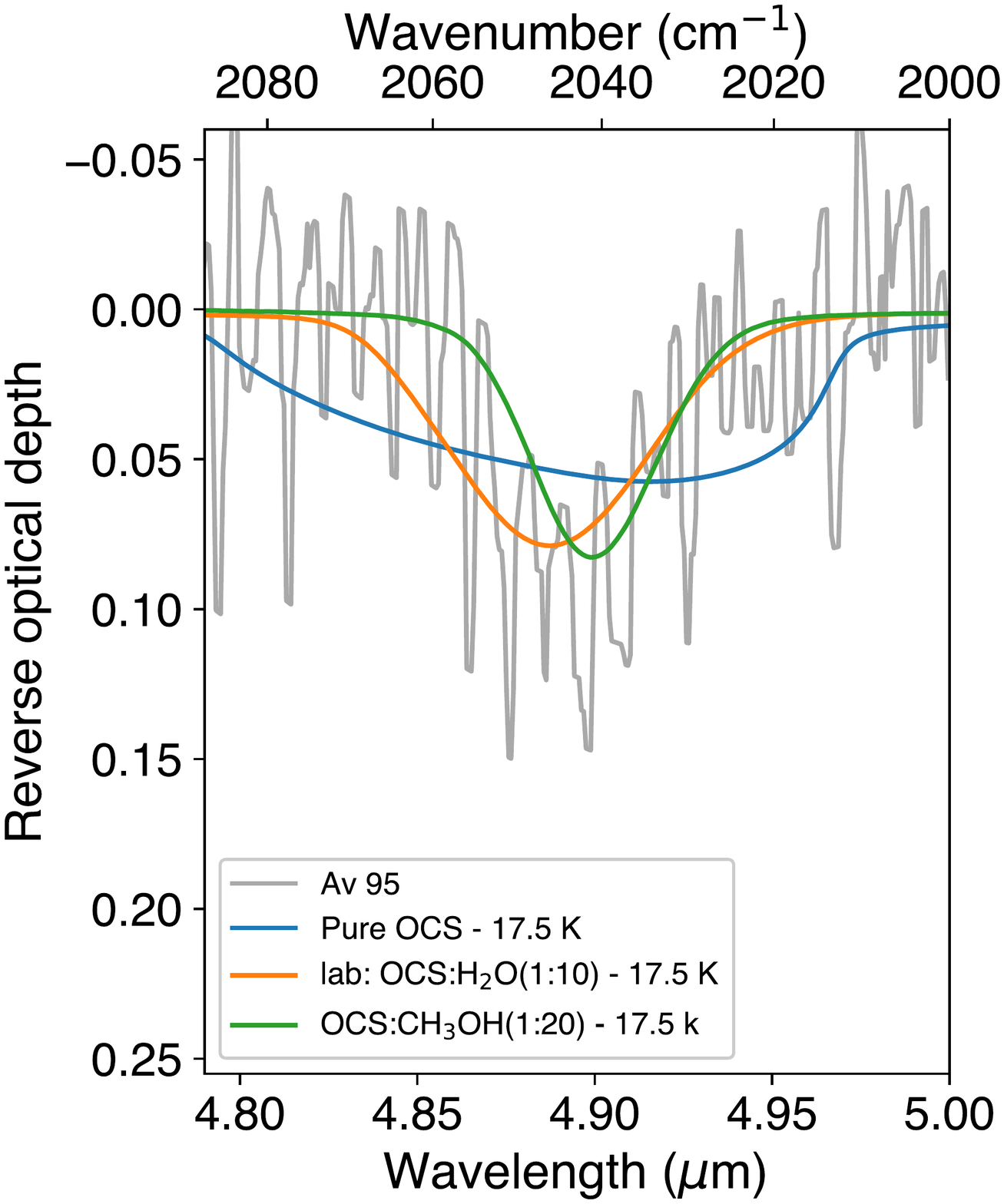}
\caption*{Extended Data Figure 8: Observed absorption profile of the C=O stretching of OCS, around 4.9 $\mu$m, in the Av = 95 source. The colored curves show the profile of the OCS in laboratory ice spectra of pure  OCS(blue), H$_2$O:OCS (orange), and CH$_3$OH:OCS (green), all at 17.5 K.}
\label{ed_fig8}
\end{figure}

\begin{figure}[htb!]
\centering
\includegraphics[totalheight=13cm]{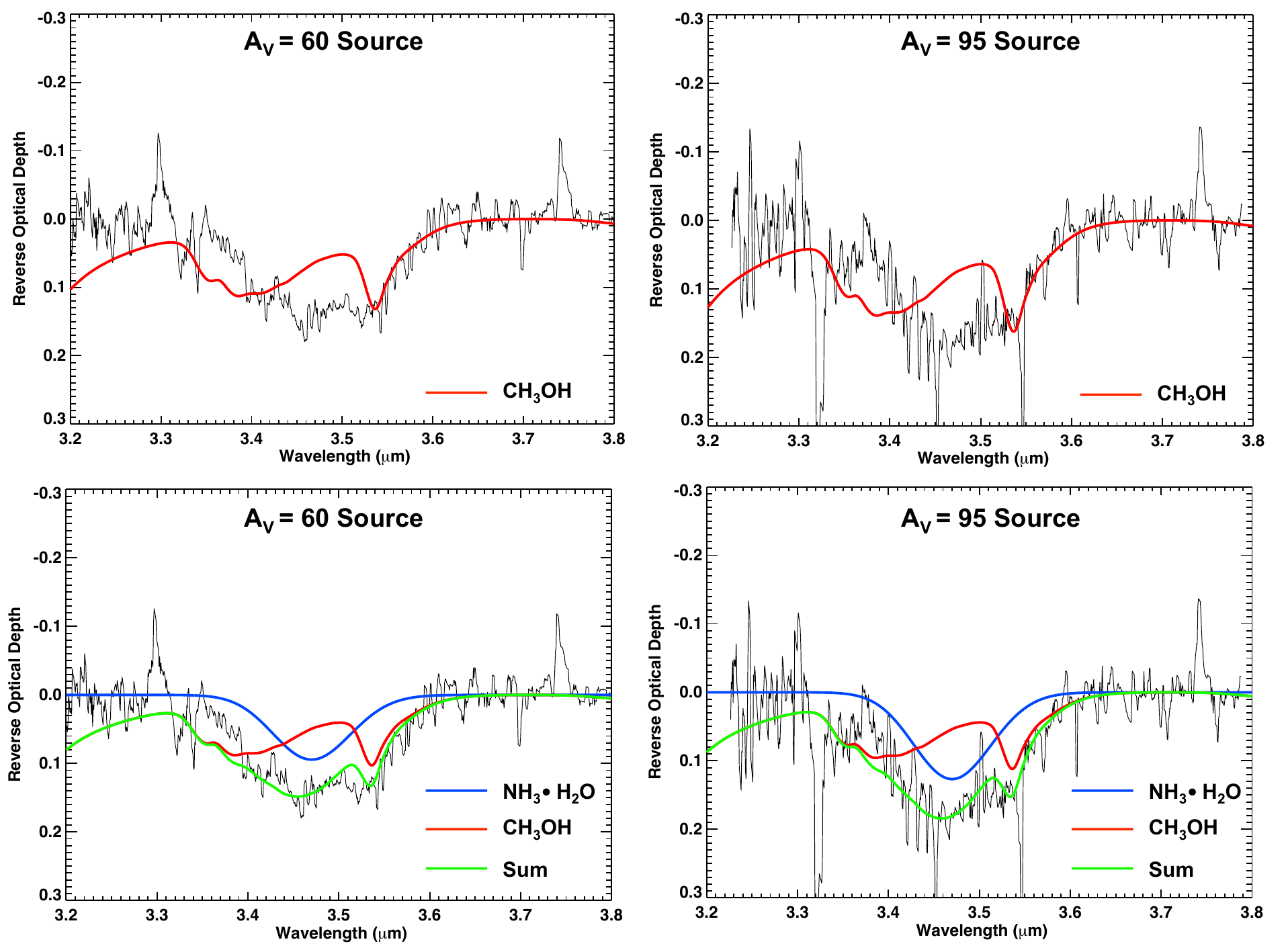}
\caption*{Extended Data Figure 9: Optical depths of the A$_V=60$ (left) and A$_V=95$ (right) background sources in the 3.2~-~3.8 $\mu$m (3125~-~2631 cm$^{-1}$) region. Top: The red line shows the optical depths of CH$_3$OH laboratory data at 15K scaled for the C-H stretching band around the 3.53 $\mu$m feature. Bottom: The blue Gaussian represents the likely {NH$_3$ $\cdot$ H$_2$O} component centered at 3.47 $\mu$m and the red line again displays the CH$_3$OH laboratory data but both are simultaneously scaled so the sum (in green) fits the data from 3.40-3.65 $\mu$m.  }

\label{ed_fig9}
\end{figure}

\begin{figure}
\resizebox{\hsize}{!}{\includegraphics{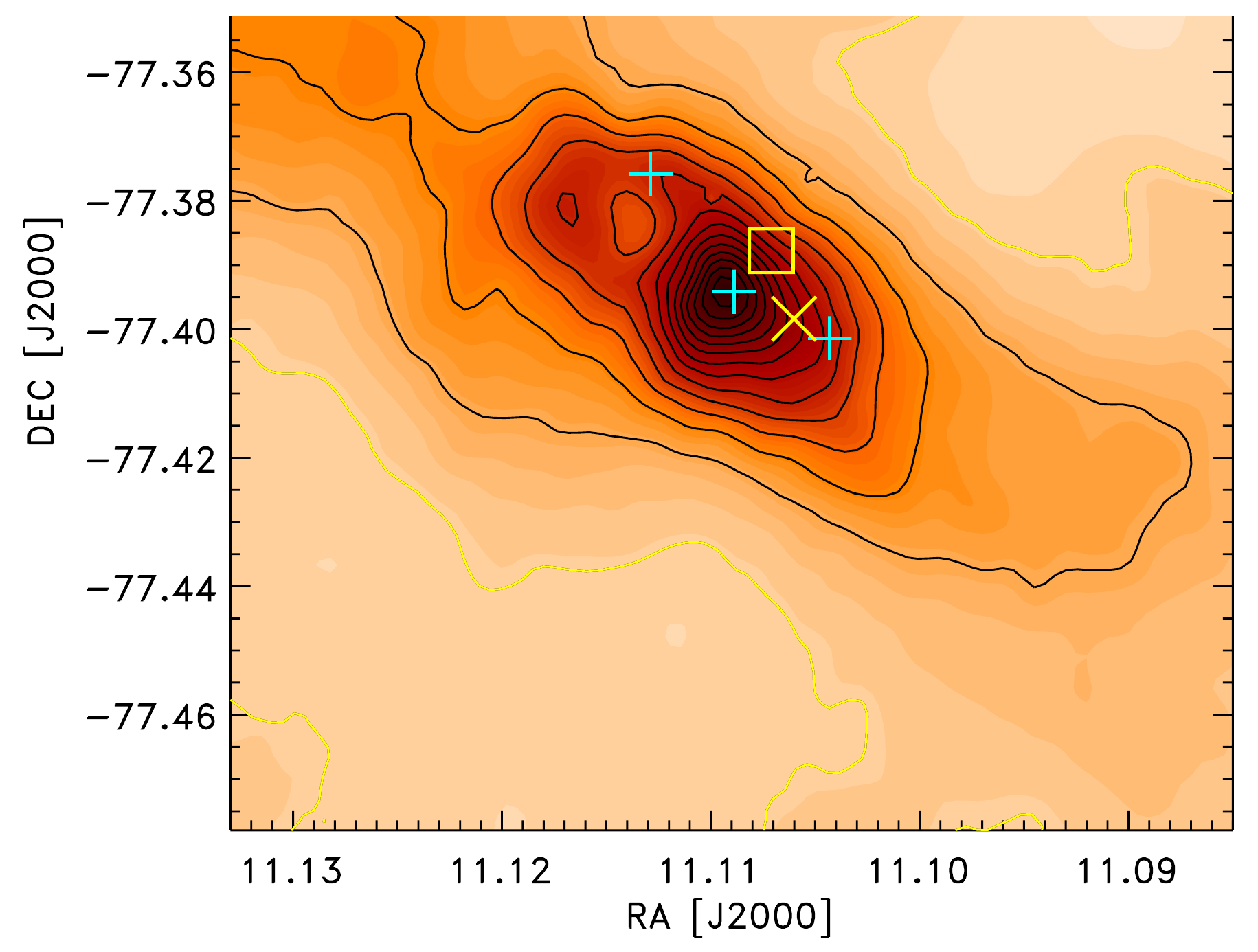}}
\caption*{Extended Data Figure 10: Map of the column density distribution in the region inferred from the \text{Herschel} far-infrared maps from 70 to 500~$\mu$m. The cyan plus-signs indicate the locations of the Class~I protostar Ced~110-IRS4, the Class~0 protostar ChamI-MMS and the clump Cha1-C2 going from the north-east (top-left) to south-west (bottom-right). The yellow box and cross indicate the location of the $A_V \approx 60$ and the $A_V \approx 95$ background stars, respectively. The contours indicate increasing H$_2$ column densities in steps of 5$\times 10^{21}$~cm$^{-2}$, starting at a value of 5$\times 10^{21}$~cm$^{-2}$ for the lowest contour (yellow line).} \label{ed_fig10}
\end{figure}

\clearpage
\section{Supplementary Information}

To validate the column densities derived from the global fitting using multiple features and mixed ice species, in Supplementary Figure 1 we show the local fits derived in the Main Text. In Supplementary Figure 2 we compare those results with the column densities determined from the local fits. The column densities from the local fits agree with those from the global fits to within the listed uncertainties. \\

\begin{figure}[htb!]
\centering
\includegraphics[width=\hsize]{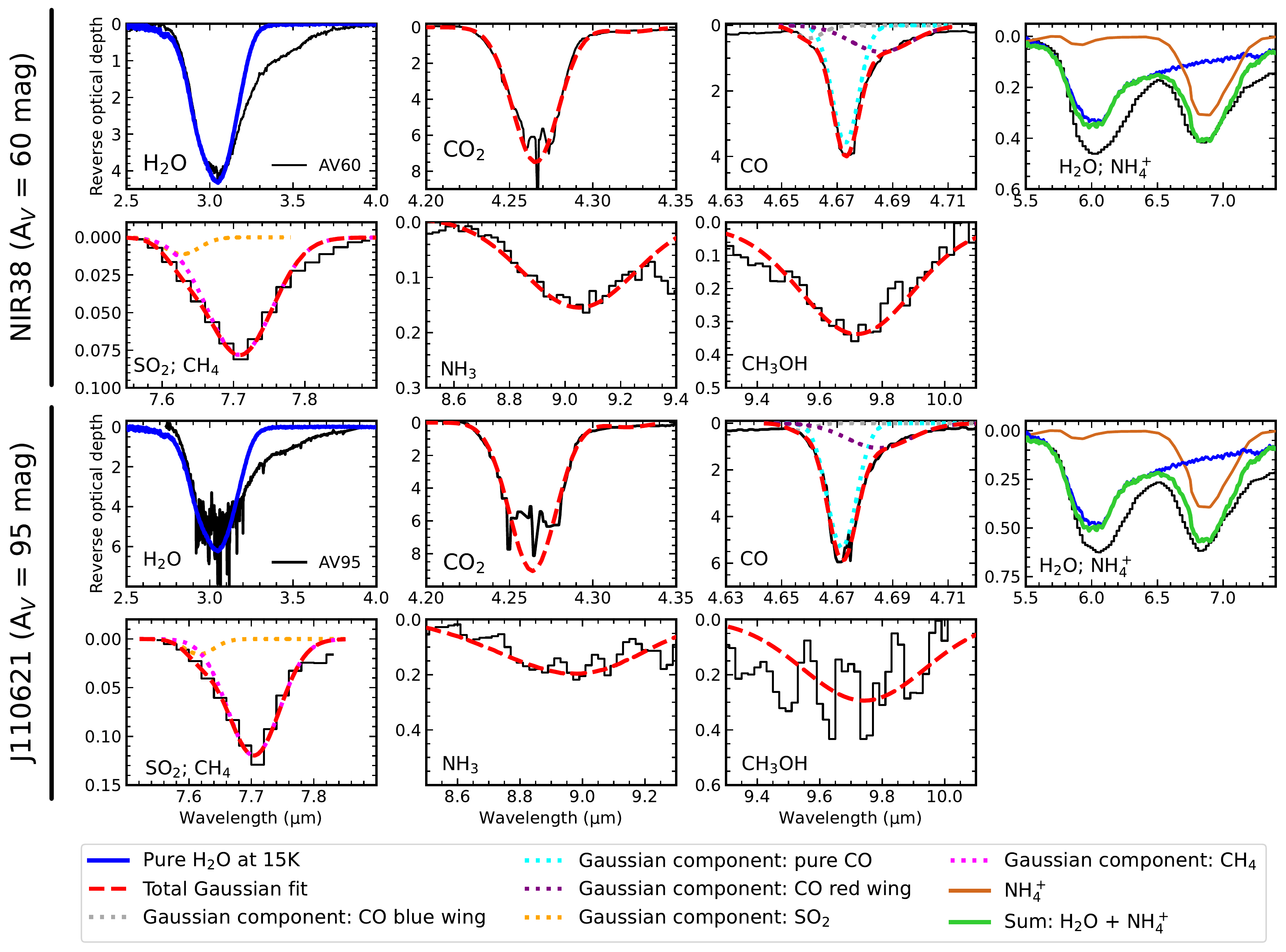}
\caption*{Supplementary Figure 1: Local fits of the major ice components in NIR38 and J110621. At 3~$\mu$m, we scale the H$_2$O ice spectrum at 15~K to match selected wavelengths that are not saturated (see Methods). Between 5.5 and 8.0~$\mu$m, we sum the NH$_4^+$ band and H$_2$O scaled to the 3~$\mu$m profile. For the other absorption bands, we perform a Gaussian fit. In the cases of CO and SO$_2$/CH$_4$, more than one Gaussian profile is adopted.}
\label{sup_fig1}
\end{figure}

\begin{figure}[htb!]
\centering
\includegraphics[width=\hsize]{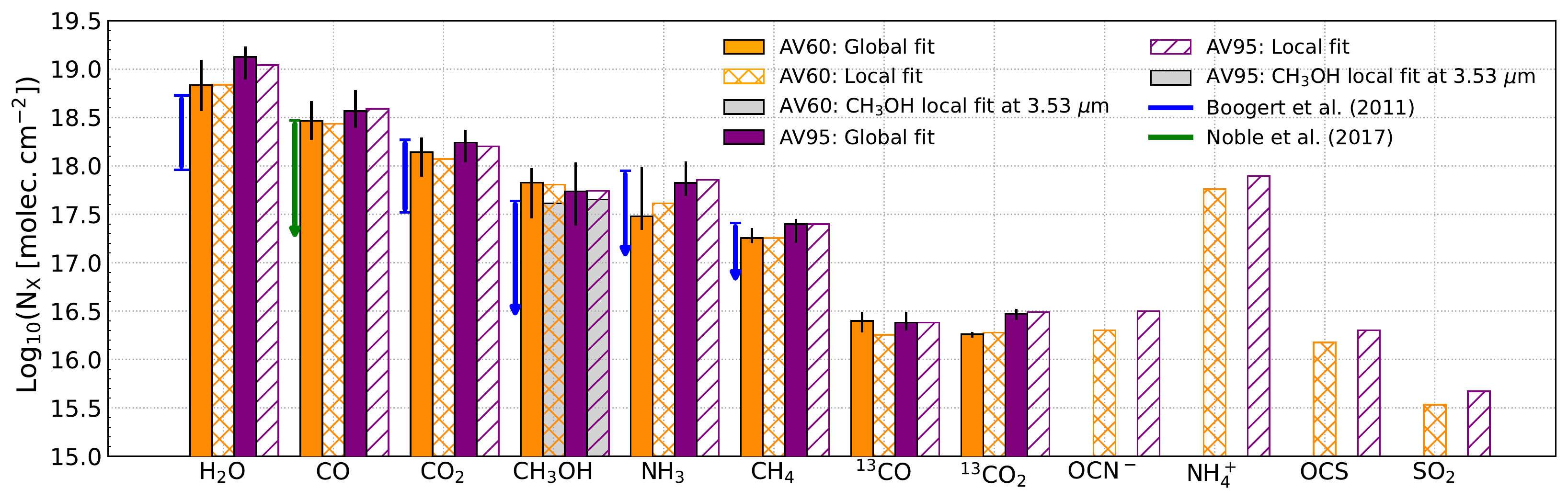}
\includegraphics[width=\hsize]{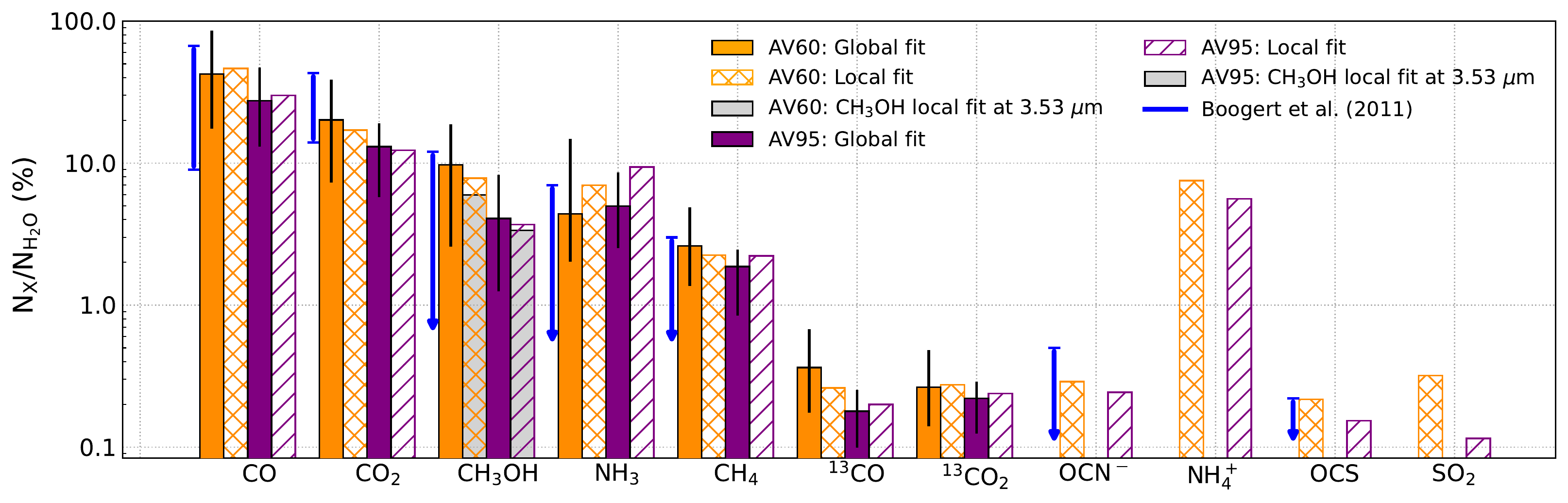}
\caption*{Supplementary Figure 2: \textbf{Barplot comparing individual column densities of ices derived from local fits to those derived from global fitting. (Top)} Ice column densities towards NIR38 ($A_V = 60$ mag) and J110621 ($A_V = 95$ mag) derived from global (full bars, best of n=112 models) and local (hatched bars) fits. Blue and green lines indicate the range of values in the literature. Arrows are used for upper limits and error bars are from the 3$\sigma$ confidence intervals. \textbf{(Bottom)} Relative column densities barplot of the detected ices, normalized to H$_2$O ice. Full and hatched bars are from global and local fits, respectively. Blue lines are from literature values, and arrows indicate upper limits. Error bars are from the 3$\sigma$ confidence intervals.}

\label{sup_fig2}
\end{figure}

\end{document}